\documentclass[12pt]{article}
\usepackage{latexsym}
\usepackage{amsmath,amsfonts}
\usepackage{times}

\catcode`\@=11

\newcount\hour
\newcount\minute
\newtoks\amorpm \hour=\time\divide\hour by 60\minute
=\time{\multiply\hour by 60 \global\advance\minute by-\hour}
\edef\standardtime{{\ifnum\hour<12 \global\amorpm={am}%
        \else\global\amorpm={pm}\advance\hour by-12 \fi
        \ifnum\hour=0 \hour=12 \fi
        \number\hour:\ifnum\minute<10
        0\fi\number\minute\the\amorpm}}
\edef\militarytime{\number\hour:\ifnum\minute<10 0\fi\number\minute}

\def\draftlabel#1{{\@bsphack\if@filesw {\let\thepage\relax
   \xdef\@gtempa{\write\@auxout{\string
      \newlabel{#1}{{\@currentlabel}{\thepage}}}}}\@gtempa
   \if@nobreak \ifvmode\nobreak\fi\fi\fi\@esphack}
        \gdef\@eqnlabel{#1}}
\def\@eqnlabel{}
\def\@vacuum{}
\def\marginnote#1{}
\def\draftmarginnote#1{\marginpar{\raggedright\scriptsize\tt#1}}
\def\draft{
        \pagestyle{plain}
        \overfullrule=2pt
        \oddsidemargin -.5truein
        \def\@oddhead{\sl \phantom{\today\quad\militarytime} \hfil
        \smash{\Large\sl DRAFT} \hfil \today\quad\militarytime}
        \let\@evenhead\@oddhead
        \let\label=\draftlabel
        \let\marginnote=\draftmarginnote
        \def\ps@empty{\let\@mkboth\@gobbletwo
        \def\@oddfoot{\hfil \smash{\Large\sl DRAFT} \hfil}
        \let\@evenfoot\@oddhead}
        \def\@eqnnum{(\theequation)\rlap{\kern\marginparsep\tt\@eqnlabel}%
        \global\let\@eqnlabel\@vacuum}  }

\newcommand{\rf}[1]{(\ref{#1})}
\renewcommand{\theequation}{\thesection.\arabic{equation}}
\renewcommand{\thefootnote}{\fnsymbol{footnote}}
\newcommand{\newsection}{   
\setcounter{equation}{0}\section}

\def\appendix#1{\addtocounter{section}{1}\setcounter{equation}{0}
\renewcommand{\thesection}{\Alph{section}}
\section*{Appendix \thesection\protect\indent \parbox[t]{11.15cm}{#1}}
\addcontentsline{toc}{section}{Appendix \thesection\ \ \ #1}}

\def\be{\begin{equation}}
\def\ee{\end{equation}}
\def\beq{\begin{eqnarray}}
\def\eeq{\end{eqnarray}}

\def\parline{\,\partial\kern -0.55em /\,\,}

\def\half{{\frac{1}{2}}}

\def\LL{{\cal L}}
\def\MM{{\cal M}}
\def\XX{{\cal X}}

\def\phik{|\phi\rangle}
\def\phibr{\langle\phi|}

\def\Phik{|\Phi\rangle}
\def\Phibr{\langle\Phi|}

\def\Xik{|\Xi\rangle}
\def\xik{|\xi\rangle}

\def\lambdak{|\lambda\rangle}

\def\tk{|t\rangle}
\def\tbr{\langle t|}

\def\smzero{{\scriptscriptstyle (0)}}
\def\smone{{\scriptscriptstyle (1)}}
\def\smtwo{{\scriptscriptstyle (2)}}

\def\smx#1{{\scriptscriptstyle (#1)}}
\def\oplussm{{\scriptscriptstyle \oplus}}
\def\ominussm{{\scriptscriptstyle \ominus}}

\def\smzero{{\scriptscriptstyle (0)}}
\def\smone{{\scriptscriptstyle (1)}}
\def\smtwo{{\scriptscriptstyle (2)}}

\def\smx#1{{\scriptscriptstyle (#1)}}
\def\oplussm{{\scriptscriptstyle \oplus}}
\def\ominussm{{\scriptscriptstyle \ominus}}

\def\rwt{\widetilde{r}}

\def\alpar{\alpha\partial}
\def\albpar{\bar\alpha\partial}

\def\eb{\bar{e}}
\def\rb{\bar{r}}

\def\nbf{{\bf n}}

\begin{document}


\begin{flushright}
FIAN-TD-2008-25 \hspace{1.2cm}{}~\\
arXiv: 0812.2861 [hep-th]\\
Modified, January 2010\hspace{0.5cm}{}~
\end{flushright}

\vspace{1cm}

\begin{center}

{\Large \bf Conformal self-dual fields }

\vspace{2.5cm}

R.R. Metsaev\footnote{ E-mail: metsaev@lpi.ru }

\vspace{1cm}

{\it Department of Theoretical Physics, P.N. Lebedev Physical
Institute, \\ Leninsky prospect 53,  Moscow 119991, Russia }

\vspace{3.5cm}

{\bf Abstract}

\end{center}

Conformal self-dual fields in flat space-time of even dimension greater than
or equal to four are studied. Ordinary-derivative formulation of such fields
is developed. Gauge invariant Lagrangian with conventional kinetic terms and
corresponding gauge transformations are obtained. Gauge symmetries are
realized by involving the Stueckelberg fields. Realization of global
conformal symmetries is obtained. Light-cone gauge Lagrangian is found. Also,
we demonstrate use of the light-cone gauge for counting of on-shell degrees
of freedom of the conformal self-dual fields.

\newpage
\renewcommand{\thefootnote}{\arabic{footnote}}
\setcounter{footnote}{0}

\section{Introduction}

In  Poincar\'e and conformal supergravity theories, the self-duality
manifests itself in different ways. In Poincar\'e supergravity theories, some
of the antisymmetric tensor fields are not self-dual, while their
field strengths are self-dual (see, e.g., Ref.\cite{Schwarz:1983wa}).%
\footnote{ It is self-duality of the field strength that leads to the problem
with Lorentz invariant action for the gauge antisymmetric tensor field
\cite{Marcus:1982yu} without the use of auxiliary fields. Study of the
Lorentz covariant formulations involving auxiliary fields may be found in
Refs.\cite{Berkovits:1996rt,Pasti:1996vs}. Interesting discussion of
self-dual fields in $d=6,10$ may be found in \cite{Bandos:2005mb}.}
In contrast to this, in conformal supergravity theories, some of the
antisymmetric tensor fields are self-dual themselves, while their field
strengths are not self-dual (see,  e.g.,  Ref.\cite{Fradkin:1985am}).

In Poincar\'e supergravity theories, the antisymmetric tensor fields are
realized as gauge fields, while in the standard approach to conformal
supergravity theories there are no gauge symmetries related to the self-dual
antisymmetric tensor fields. Note also that the antisymmetric tensor fields
of Poincar\'e supergravity theories describe ghost-free dynamics, while the
ones of conformal supergravity theories contain  of ghost degrees of freedom.

In this paper we discuss   the self-dual antisymmetric tensor fields of
conformal supergravity theories (which are well defined  only   in $d=4,6$)
and their counterparts in space-times of arbitrary even dimensions. It is
these self-dual antisymmetric tensor fields that will be referred to as
conformal self-dual fields, or shortly as self-dual fields in this paper.

The standard formulation of the self-dual fields
involves exotic kinetic terms%
\footnote{For instance, the self-dual field $T^{ab}$ of ${\cal N}=4$, $4d$
conformal supergravity is described by the Lagrangian $\LL=
\partial^a\bar{T}^{ab}\partial^c T^{cb}$.}.
These exotic kinetic terms can be reexpressed in terms of kinetic terms
involving the standard Dalambertian operator but this leads to higher
derivatives (see e.g. Ref.\cite{Fradkin:1985am}). Also, as was mentioned
above, in the standard approach, there are no gauge symmetries associated  to
such self-dual fields.

The purpose of this paper is to develop an ordinary (not higher-) derivative,
gauge invariant,
and Lagrangian formulation for the self-dual fields%
\footnote{ Making comparison with various approaches to massive fields, one
can say that the standard approach to the self-dual fields is a counterpart
of the Pauli-Fierz approach to the massive fields, while our approach to the
self-dual fields is a counterpart of the Stueckelberg approach to the massive
fields.}.
In this paper, we discuss free self-dual fields in space-time of even
dimension $d \geq 4$. Our approach to the self-dual fields can be summarized
as follows.

i) We introduce additional field degrees of freedom (D.o.F), i.e., we extend
space of fields entering the standard formulation of self-dual fields. These
additional field D.o.F are supplemented by appropriate gauge symmetries%
\footnote{ To realize those additional gauge symmetries we adopt the approach
of Refs.\cite{Zinoviev:2001dt,Metsaev:2006zy} which turns out to be the most
useful for our purposes.}.
We note that these additional field D.o.F are similar to the ones used in the
gauge invariant formulation of massive fields. Sometimes, such additional
field D.o.F are referred to as Stueckelberg fields.

ii) Our Lagrangian for the free self-dual fields does not involve higher than
second order terms in derivatives. Two-derivative contributions to the
Lagrangian take the form of the standard kinetic terms of the antisymmetric
tensor fields. The Lagrangian is invariant under gauge transformations and
global conformal algebra transformations.

iii) Gauge transformations of the free self-dual fields do not involve higher
than first order terms in derivatives. One-derivative contributions to the
gauge transformations take the form of the standard gauge transformations of
the antisymmetric tensor fields.

iv) The gauge symmetries of our Lagrangian make it possible to match  our
approach with  the standard one, i.e., by an appropriate gauge fixing of the
Stueckelberg fields and by solving some constraints we obtain the standard
formulation of the self-dual fields. This implies that our approach retain
on-shell D.o.F of the standard theory of self-dual fields, i.e., on-shell,
our approach is equivalent to the standard one.

As is well known, the Stueckelberg approach turned out to be successful for
the study of theories involving massive fields. That  is to say that all
covariant formulations of string theories are realized by using Stueckelberg
gauge symmetries. The self-dual fields enter field content of conformal
supergravity theories. Therefore we expect that use of the Stueckelberg
fields for the studying self-dual fields might be useful for developing new
interesting formulations of the conformal supergravity theories.

The rest of the paper  is organized as follows.

In Sec. \ref{sec02},  we summarize the notation and review the standard
approach to the self-dual fields.

In Sec. \ref{man02-sec-03}, we  start with  the example of self-dual field
propagating in $4d$ Minkowski space. For this field, we obtain the
ordinary-derivative gauge invariant Lagrangian. We find realization of the
conformal $so(4,2)$ algebra symmetries on space of gauge fields and on space
of field strengths. Also we obtain light-cone gauge Lagrangian and
demonstrate that number of on-shell D.o.F of our approach coincides with the
one in the standard approach to the self-dual field.  We discuss the
decomposition of those on-shell D.o.F into irreps of the $so(2)$algebra.

In Sec. \ref{man02-sec-04}, we generalize results obtained in Sec.
\ref{man02-sec-03} to the case of self-dual fields propagating in Minkowski
space of arbitrary dimension.

In Sec. \ref{man02-sec-05}, we represent our results in Secs.
\ref{man02-sec-03}, \ref{man02-sec-04} by using realization of field degrees
of freedom in terms of generating functions. The generating functions are
constructed out of the self-dual gauge fields and some oscillators. Use of
the generating functions simplifies considerable study of the self-dual
fields. Therefore we believe that result in Sec. \ref{man02-sec-05} might be
helpful in future studies of the self-dual fields.

Section \ref{conl-sec-01} suggests directions for future research.

We collect various technical details in appendices. In Appendix A, we discuss
details of the derivation of the ordinary-derivative gauge invariant
Lagrangian. In Appendix B, we present details of the derivation of the
conformal algebra transformations of gauge fields. In Appendix C, we discuss
some details of the derivation of the light-cone gauge Lagrangian. In
Appendix D we collect some useful formulas involving the Levi-Civita symbol.

\newsection{Preliminaries}\label{sec02}

\subsection{Notation}

Throughout the paper, dimension of a flat space-time, which we denote by $d$,
is restricted to be even integer, $d=2\nu$.  Coordinates in the flat
space-time are denoted by $x^a$, while $\partial_a$ stands for derivative
with respect to $x^a$, $\partial_a \equiv \partial / \partial x^a$. Vector
indices of the Lorentz algebra $so(d-1,1)$ take the values
$a,b,c,e=0,1,\ldots ,d-1$. To simplify our expressions we drop $\eta_{ab}$ in
scalar products, i.e. we use $X^aY^a \equiv \eta_{ab}X^a Y^b$. The notation
$\epsilon^{a_1\ldots a_\nu b_1 \ldots b_\nu}$ stands for the Levi-Civita
symbol. We assume the normalization $\epsilon^{01\ldots d-1}=1$.

To avoid complicated tensor expressions we use a set of the creation
operators $\alpha^a$, $\zeta$, $\upsilon^\oplussm$, $\upsilon^\ominussm$, and
the respective set of annihilation operators $\bar{\alpha}^a$, $\bar{\zeta}$,
$\bar\upsilon^\ominussm$, $\bar\upsilon^\oplussm$,
\be \bar\alpha^a |0\rangle = 0\,,\qquad  \bar\zeta|0\rangle = 0\,,\qquad
\bar\upsilon^\oplussm |0\rangle = 0\,,\qquad \bar\upsilon^\ominussm |0\rangle
= 0\,.\ee
These operators satisfy the following (anti)commutation relations:
\beq
& & \{\bar{\alpha}^a,\alpha^b\}=\eta^{ab}\,, \qquad \{\bar\zeta,\zeta\}=1\,,
\\
& & [\bar{\upsilon}^\oplussm,\, \upsilon^\ominussm ]=1\,, \qquad\quad
[\bar{\upsilon}^\ominussm,\, \upsilon^\oplussm]=1\,,
\eeq
and will often be referred to as oscillators in what follows%
\footnote{ We use oscillator formulation
\cite{Lopatin:1987hz}-\cite{Bekaert:2006ix} to handle the many indices
appearing for tensor fields. It can also be reformulated as an algebra acting
on the symmetric-spinor bundle on the manifold $M$ \cite{Hallowell:2005np}.
Note that the scalar oscillators $\zeta$, $\bar\zeta$ arise naturally by a
dimensional reduction from flat space. It is natural to expect that the
`conformal' oscillators $\upsilon^\oplussm$, $\upsilon^\ominussm$,
$\bar\upsilon^\oplussm$, $\bar\upsilon^\ominussm$ also allow certain
interpretation via dimensional reduction. Interesting recent discussion of
dimensional reduction may be found in Ref.\cite{Artsukevich:2008vy}.}.
The oscillators $\alpha^a$, $\bar\alpha^a$ and $\zeta$, $\bar\zeta$,
$\upsilon^\oplussm$, $\upsilon^\ominussm$, $\bar\upsilon^\oplussm$,
$\bar\upsilon^\ominussm$ transform in the respective vector and scalar
representations of the $so(d-1,1)$ Lorentz algebra and satisfy the following
hermitian conjugation rules:
\beq
\alpha^{a\dagger} = \bar\alpha^a\,, \qquad \zeta^\dagger = \bar\zeta \,,
\qquad
\upsilon^{\oplussm\dagger} = \bar\upsilon^\oplussm\,,\qquad
\upsilon^{\ominussm \dagger} = \bar\upsilon^\ominussm \,.
\eeq
Throughout this paper we use operators constructed out of the oscillators and
derivatives,
\be \Box=\partial^a\partial^a\,,\qquad  \alpha\partial =\alpha^a\partial^a\,,
\qquad \bar\alpha\partial =\bar\alpha^a\partial^a\,,\ee
\beq \label{10122208-02}
&& N_\alpha \equiv \alpha^a \bar\alpha^a \,,
\\
&& N_\zeta \equiv \zeta \bar\zeta \,,
\\
&& N_{\upsilon^\oplussm} \equiv \upsilon^\oplussm
\bar\upsilon^\ominussm\,,
\\
&& N_{\upsilon^\ominussm} \equiv \upsilon^\ominussm
\bar\upsilon^\oplussm\,,
\\
\label{10122208-07} && N_\upsilon \equiv N_{\upsilon^\oplussm} +
N_{\upsilon^\ominussm} \,.
\eeq

\subsection{Global conformal symmetries }

In space-time of dimension $d$, the conformal algebra $so(d,2)$ referred to
the basis of Lorentz algebra $so(d-1,1)$ consists of translation generators
$P^a$, conformal boost generators $K^a$, dilatation generator $D$ and
generators $so(d-1,1)$ Lorentz algebra $J^{ab}$. We assume the following
normalization for commutators of the conformal algebra:
\beq
\label{ppkk}
&& {}[D,P^a]=-P^a\,, \hspace{2cm}  {}[P^a,J^{bc}]=\eta^{ab}P^c -\eta^{ac}P^b
\,,
\\
\label{dppkk} && [D,K^a]=K^a\,, \hspace{2.2cm} [K^a,J^{bc}]=\eta^{ab}K^c -
\eta^{ac}K^b\,,
\\[5pt]
\label{pkjj} && \hspace{2.5cm} {}[P^a,K^b]=\eta^{ab}D-J^{ab}\,,
\\
&& \hspace{2.5cm} [J^{ab},J^{ce}]=\eta^{bc}J^{ae}+3\hbox{ terms} \,.
\eeq

Let $\phik$ denotes field propagating in flat space-time of dimension $d\geq
4$. Let Lagrangian for the free field $\phik$ be conformal invariant. This
implies, that Lagrangian is invariant with respect to transformation
(invariance of the Lagrangian is assumed to be up to total derivative)
\be \label{man-12112009-03} \delta_{\hat{G}} \phik  = \hat{G} \phik \,, \ee
where a realization of the conformal algebra generators $\hat{G}$ in terms of
differential operators takes the form
\beq
\label{conalggenlis01} && P^a = \partial^a \,,
\\[3pt]
\label{conalggenlis02} && J^{ab} = x^a\partial^b -  x^b\partial^a + M^{ab}\,,
\\[3pt]
\label{conalggenlis03} && D = x\partial  + \Delta\,,
\\[3pt]
\label{conalggenlis04} && K^a = K_{\Delta,M}^a + R^a\,,
\\[3pt]
\label{conalggenlis05} &&\qquad  K_{\Delta,M}^a \equiv
-\frac{1}{2}x^2\partial^a + x^a D + M^{ab}x^b\,,
\\[3pt]
&& \qquad x\partial \equiv x^a \partial^a \,, \qquad x^2 = x^a x^a\,.\eeq
In \rf{conalggenlis02}-\rf{conalggenlis04}, $\Delta$ is operator of conformal
dimension, $M^{ab}$ is spin operator of the Lorentz algebra,
\be  [M^{ab},M^{ce}]=\eta^{bc}M^{ae}+3\hbox{ terms} \,, \ee
and $R^a$ is operator depending on the derivative $\partial^a$ and not
depending on the space-time coordinates $x^a$, $[P^a,R^b]=0$. The spin
operator $M^{ab}$ is well known for arbitrary tensor fields of the Lorentz
algebra. In the standard formulation of the self-dual fields, the operator
$R^a$ is equal to zero, while in the ordinary-derivative approach, we develop
in this paper, the operator $R^a$ is non-trivial. This implies that, in the
framework of ordinary-derivative approach, the complete description of the
self-dual fields requires finding not only gauge invariant Lagrangian but
also the operator $R^a$.

Explicit representation for the action of operator $K_{\Delta,M}^a$
\rf{conalggenlis05} is easily obtained from the relations above-given. Let
$\Lambda^{a_1\ldots a_n}$ be rank-$n$ antisymmetric tensor field of the
Lorentz algebra $so(d-1,1)$, while $\Delta(\Lambda)$ is a conformal dimension
of this tensor field. Relation \rf{conalggenlis04} implies that the conformal
boost transformations of $\Lambda^{a_1\ldots a_n}$ can be presented as
\beq \label{conalggenlis04xxx}
\delta _{K^a} \Lambda^{a_1\ldots a_n} & = &  \delta _{K_{\Delta,M}^a}
\Lambda^{a_1\ldots a_n} + \delta _{R^a} \Lambda^{a_1\ldots a_n} \,,
\\
\label{conalggenlis04xxxnew} && \delta _{K_{\Delta,M}^a} \Lambda^{a_1\ldots
a_n} = K_{\Delta(\Lambda)}^a \Lambda^{a_1\ldots a_n} + \sum_{k=1}^n M^{aa_k c
} \Lambda^{a_1\ldots a_{k-1} c a_{k+1}\ldots a_n} \,,
\\
\label{man10112009-04} &&  K_{\Delta}^a \equiv -\half x^2 \partial^a + x^a
(x\partial + \Delta) \,,
\\
&&  M^{abc} \equiv \eta^{ab}x ^c -\eta^{ac} x^b \,. \eeq
Thus, all that remains is to find explicit representation for the operator
$R^a$. This is what we are doing, among other things, in this paper.

\subsection{ Standard approach to self-dual fields}

We begin with brief review of the standard approach to the conformal
self-dual fields. In this section we recall main facts of conformal field
theory about these fields.

Consider totally antisymmetric rank-$\nu$ tensor field $T^{a_1\ldots a_\nu}$,
$\nu \equiv \frac{d}{2}$, of the Lorentz algebra $so(d-1,1)$, where the
dimension of space-time $d$ is even integer. In the framework of standard
approach, the field $T^{a_1\ldots a_\nu}$ is referred to as conformal
self-dual field if it meets the following requirements:

{\bf a}) The field $T^{a_1\ldots a_\nu}$ satisfies the self-duality
constraint

\be \label{08122008-01}  T^{a_1\ldots a_\nu } =
\frac{\tau}{\nu!}\epsilon^{a_1\ldots a_\nu b_1\ldots b_\nu }T^{b_1\ldots
b_\nu}\,,\ee

\be \label{man10112009-01} \tau =  \left\{\begin{array}{l}
\pm {\rm i} \ \ \hbox{ for } \ \ d = 4k;
\\[7pt]
\pm 1 \ \ \hbox{ for } \ \ d = 4k+2;
\end{array}\right.
\ee
where $\epsilon^{a_1\ldots a_\nu b_1\ldots b_\nu }$ is the Levi-Civita
symbol. For flexibility, we do not fix sign of $\tau$. Constraint
\rf{08122008-01} implies that $T^{a_1\ldots a_\nu}$ is {\it complex-valued}
when $d=4k$. In $d=4k+2$ the field $T^{a_1\ldots a_\nu}$ is considered to be
{\it real-valued}.

{\bf b}) Dynamics of the field $T^{a_1\ldots a_\nu}$ is described by the
Lagrangian
\beq
\label{04122008-01} && \LL_{\rm st} = \frac{1}{(\nu-1)!} \partial^a
\bar{T}^{aa_2\ldots a_\nu }
\partial^b T^{ba_2\ldots a_\nu } \,, \qquad \hbox{ for } d=4k\,,
\\[5pt]
\label{04122008-01xx} && \LL_{\rm st} = \frac{1}{(\nu-1)!} \partial^a
T^{aa_2\ldots a_\nu } \partial^b T^{ba_2\ldots a_\nu } \,, \qquad \hbox{ for
} d=4k+2\,,
\eeq
where $\bar{T}$ in \rf{04122008-01} stands for complex conjugate of $T$.

We now note that:
\\
{\bf i}) Requiring the Lagrangian to be invariant under the dilatation
transformation we obtain the conformal dimension of the field $T^{a_1\ldots
a_\nu}$,
\\
\be\label{condim0001} \Delta(T^{a_1\ldots a_\nu}) = \frac{d-2}{2}\,, \ee
which is referred to as {\it the canonical conformal dimension of the
conformal self-dual field}.\\
{\bf ii}) Operator $R^a$ of the field $T^{a_1\ldots a_\nu}$ is equal to zero.
\\
{\bf iii}) Simplest case of self-dual field, which is antisymmetric
complex-valued rank-2 tensor field $T^{ab}$, corresponds to $d=4$ with the
following self-duality constraint (see \rf{08122008-01})
\be T^{ab} = \frac{\tau }{2}\epsilon^{abce}T^{ce}\,.\ee
The Lagrangian for the field $T^{ab}$ can be read from \rf{04122008-01}
\be \label{standlag01} \LL_{\rm st} = \partial^a \bar{T}^{ac} \partial^b
T^{bc}\,.\ee
The field $T^{ab}$ appears in the field content of ${\cal N}=4$, $4d$
conformal supergravity.

\newsection{ Ordinary-derivative approach to
self-dual field for $d=4$} \label{man02-sec-03}

As a warm up let us start with the simplest case of the self-dual fields.
Consider the self-dual field $T^{ab}$ propagating in $4d$ flat space. In the
framework of ordinary-derivative approach, a dynamical system that on-shell
equivalent to the self-dual field $T^{ab}$ with Lagrangian \rf{standlag01}
involves two vector fields $\phi_{-1}^a$ and $\phi_1^a$, one scalar field
$\phi_0$ and one self-dual rank-2 tensor field $t^{ab}$. In other words, we
use the following field content:
\be \label{scafiecol01}
\phi_{-1}^a\,,\qquad \phi_1^a\,,\qquad \phi_0\,, \qquad t^{ab}\,,
\ee
\be \label{08122208-02} t^{ab} = \frac{\tau}{2}\epsilon^{abce}t^{ce}\,,\ee
$\tau = \pm {\rm i}$. All fields in \rf{scafiecol01} are {\it
complex-valued}. Conformal dimensions of these fields are given by
\be \label{delscalcomdef01}
\Delta(\phi_{-1}^a) = 0\,,\qquad \Delta(\phi_1^a) =2\,,\qquad
\Delta(\phi_0)=1\,, \qquad \Delta(t^{ab})= 1\,.
\ee
We note that subscript $k'$ in $\phi_{k'}$ implies that conformal dimension
of the field $\phi_{k'}$ is equal to $1+k'$.

Ordinary-derivative action and Lagrangian  we found take the form

\be S = \int d^4x \, \LL \,, \ee

\beq \label{080312-23}
\LL &= & -\half F^{ab}(\bar\phi_{-1}) F^{ab}(\phi_1) -\half
F^{ab}(\bar\phi_1) F^{ab}(\phi_{-1})
\nonumber\\[5pt]
&- & \half \bar{t}^{ab} F^{ab}(\phi_1)-\half F^{ab}(\bar\phi_1) t^{ab}
- (\bar\phi_1^a + \partial^a \bar\phi_0)(\phi_1^a + \partial^a \phi_0)\,,
\eeq
where $\bar\phi_{k'}$ and $\bar{t}^{ab}$ are the respective complex
conjugates of $\phi_{k'}$ and $t^{ab}$, while $F^{ab}(\phi)$ stands for field
strength defined as
\be \label{04122208-02}  F^{ab}(\phi) \equiv \partial^a \phi^b -
\partial^b \phi^a\,. \ee
Details of the derivation of Lagrangian \rf{080312-23} may be found in
Appendix A.

A few remarks are in order.\\
{\bf i)} Two-derivative contributions to Lagrangian \rf{080312-23} are the
standard kinetic terms for the vector fields $\phi_{-1}^a$, $\phi_1^a$ and
the standard Klein-Gordon kinetic term for the scalar field $\phi_0$.
\\
{\bf ii}) In addition to the two-derivative contributions, the Lagrangian
involves one-derivative contributions and derivative-independent mass-like
contributions. Appearance of the one-derivative and derivative-independent
contributions to the Lagrangian is a characteristic feature of the
ordinary-derivative approach.
\\
{\bf iii}) The self-dual field $t^{ab}$ plays the role of a Lagrangian
multiplier. Equations of motion for $t^{ab}$ together with self-duality
constraint for $t^{ab}$ \rf{08122208-02} tell us that {\it on-shell} the
field strength $F^{ab}(\phi_1)$ is self-dual,
\be
F^{ab}(\phi_1) = \frac{\tau }{2} \epsilon^{abce} F^{ce}(\phi_1)\,.
\ee

{\it Gauge transformations}. To discuss gauge symmetries of Lagrangian
\rf{080312-23} we introduce the following gauge transformation parameters:

\be \xi_{-2}\,,\qquad \xi_0\,,\qquad \lambda^a \,. \ee
All these gauge transformation parameters are complex-valued. Conformal
dimensions of the gauge transformation parameters are given by
\be
\Delta(\xi_{-2}) = -1\,,\qquad \Delta(\xi_0) =1\,,\qquad \Delta(\lambda^a)=
0\,.
\ee

We find that Lagrangian \rf{080312-23} is invariant under the gauge
transformations
\beq
\label{080312-24} && \delta\phi_1^a = \partial^a \xi_0 \,,
\\
\label{080312-25} && \delta\phi_{-1}^a = \partial^a \xi_{-2} - \lambda^a \,,
\\
\label{080312-26} && \delta\phi_0 = - \xi_0 \,,
\\
\label{080312-27} && \delta t^{ab} = F^{ab}(\lambda) + \frac{\tau}{2}
\epsilon^{abce} F^{ce}(\lambda)\,,
\eeq
where field strength $F^{ab}(\lambda)$ for the gauge transformation parameter
$\lambda^a$ is defined as in \rf{04122208-02}.

From \rf{080312-25},\rf{080312-26}, we see that the vector and scalar fields,
$\phi_{-1}^a$, $\phi_0$, transform as Stueckelberg fields, i.e., these fields
can be gauged away via Stueckelberg gauge fixing, $\phi_{-1}^a=0$,
$\phi_0=0$. If we gauge away these fields, and exclude the vector field
$\phi_1^a$ via equations of motion, then our Lagrangian reduces to the
Lagrangian of standard approach \rf{standlag01}. Note also that, in the
Stueckelberg gauge, the field $t^{ab}$ is identified with the generic
self-dual field $T^{ab}$.  Thus, our approach is equivalent to the standard
one.

{\it Realization of conformal algebra symmetries}. To complete the
ordinary-derivative description of the conformal self-dual field we should
provide realization of the conformal algebra symmetries on a space of fields
\rf{scafiecol01}. The Poincar\'e algebra symmetries are realized on fields
\rf{scafiecol01} in a standard way. Realization of dilatation symmetry is
given by \rf{conalggenlis03}, where conformal dimensions of fields
\rf{scafiecol01} are given in \rf{delscalcomdef01}. General realization of
conformal boost symmetries on arbitrary tensor fields is given in
\rf{conalggenlis04xxx}. According to \rf{conalggenlis04xxx} we should find
realizations of the operators $K_{\Delta,M}^a$ and $R^a$ on a space of fields
\rf{scafiecol01}. The realization of the former operator is obtained by
adopting general formula \rf{conalggenlis04xxxnew} for gauge fields
\rf{scafiecol01},

\beq \label{man10112009-05}
&& \delta_{K_{\Delta,M}^a}\phi_{k'}^b = K_{\Delta(\phi_{k'})}^a\phi_{k'}^b +
M^{abc} \phi_{k'}^c\,, \qquad k' =\pm1\,,
\\[3pt]
\label{man10112009-06} && \delta_{K_{\Delta,M}^a}\phi_0 =
K_{\Delta(\phi_0)}^a\phi_0\,,
\\[3pt]
\label{man10112009-07} && \delta_{K_{\Delta,M}^a}t^{a_1a_2} = K_{\Delta(t)}^a
t^{a_1a_2} + M^{aa_1c} t^{ca_2}+ M^{aa_2c} t^{a_1c}\,,
\eeq
where the operator $K_\Delta^a$ and conformal dimensions are defined in
\rf{man10112009-04} and \rf{delscalcomdef01} respectively.

The real difficulty is to find the operator $R^a$. Realization of the
operator $R^a$ on a space of gauge fields \rf{scafiecol01} we found is given
by

\beq \label{man03-13082009-01}
&& \delta_{R^a} \phi_1^b  = -t^{ab} - \eta^{ab} \phi_0
-  \partial^a \phi_{-1}^b - \frac{\tau}{2}\epsilon^{abce}F^{ce}(\phi_{-1})\,,
\\[5pt]
\label{man03-13082009-02} && \delta_{R^a} \phi_{-1}^b  =  0\,,
\\[5pt]
\label{man03-13082009-03} && \delta_{R^a} \phi_0  =  \phi_{-1}^a\,,
\\[5pt]
\label{man03-13082009-04} && \delta_{R^a} t^{a_1 a_2} = \eta^{a a_1}
\phi_{-1}^{a_2} -
\eta^{aa_2} \phi_{-1}^{a_1}
+ \tau \epsilon^{a_1a_2 a b } \phi_{-1}^b\,.
\eeq
Using \rf{man10112009-05}-\rf{man03-13082009-04} and general formula
\rf{conalggenlis04xxx},
\be \delta_{K^a} =\delta_{K_{\Delta,M}^a}+ \delta_{R^a}\,, \ee
gives the conformal boost transformations of the gauge fields.

From \rf{man03-13082009-01}-\rf{man03-13082009-04}, we see the operator $R^a$
maps the gauge field with conformal dimension $\Delta$ into the ones having
conformal dimension less than $\Delta$. This is to say that the realization
of the operator $R^a$ given in \rf{man03-13082009-01}-\rf{man03-13082009-04}
can schematically be represented as
\beq
&  \phi_1 \stackrel{R}{\longrightarrow} t\oplus \phi_0 \oplus
\partial \phi_{-1}\,, &
\\[7pt]
& t \stackrel{R}{\longrightarrow} \phi_{-1}\,,\qquad \phi_0
\stackrel{R}{\longrightarrow} \phi_{-1}\,,  \qquad
\phi_{-1}\stackrel{R}{\longrightarrow} 0\,. &  \eeq
Details of the derivation of the operator $R^a$ may be found in Appendix B.%
\footnote{ In appendix B, the operator $R^a$ is obtained by using the
realization of field D.o.F in terms of generating functions. Discussion of
the generating functions may be found in Sec. \ref{man02-sec-05}. Therefore,
before reading Appendix B, the reader should consult Sec.
\ref{man02-sec-05}.}
As a side of remak we note that, having introduced field content and the
Lagrangian, the operator $R^a$ is fixed uniquely by requiring that\\
{\bf i}) the operator $R^a$ should not involve higher than first order terms
in the derivative;\\
{\bf ii}) Lagrangian should be invariant under the conformal algebra
transformations.

As usually, the conformal algebra transformations of gauge fields
\rf{scafiecol01} are defined up to gauge transformations. Alternatively, the
conformal algebra symmetries can be realized on a space of field strengths.
We now discuss field strengths for gauge fields \rf{scafiecol01} and the
corresponding conformal transformations of the field strengths.

{\it Realization of conformal algebra symmetries on space of field
strengths}. We introduce the following field strengths which are constructed
out of gauge fields \rf{scafiecol01}:
\beq \label{man03-13082009-05}
&& F^{\ominussm ab} = T^{ab}\,,
\\[5pt]
\label{man03-13082009-06} && F^{abc} = F^{abc}(T)\,,
\\[5pt]
\label{man03-13082009-07} && F^{\oplussm\ominussm a} = \phi_1^a +
\partial^a\phi_0\,,
\\[5pt]
\label{man03-13082009-08} && F^{\oplussm ab} = F^{ab}(\phi_1)\,,
\eeq
where $F^{ab}(\phi_1)$ is defined as in \rf{04122208-02}, while  $T^{ab}$ and
$F^{abc}(T)$ are defined by the respective relations
\beq
\label{man03-13082009-10} && T^{ab} = t^{ab} + F^{ab}(\phi_{-1}) +
\frac{\tau}{2}\epsilon^{abce}F^{ce}(\phi_{-1})\,,
\\
\label{man03-13082009-09} && F^{abc}(T) = \partial^a T^{bc} + \partial^b
T^{ca} + \partial^c T^{ab}\,.
\eeq
One can make sure that field strengths
\rf{man03-13082009-05}-\rf{man03-13082009-08} are invariant under gauge
transformations \rf{080312-24}-\rf{080312-27}. Conformal dimensions of the
field strengths can easily be read from relations \rf{delscalcomdef01} and
\rf{man03-13082009-05}-\rf{man03-13082009-08}
\be \label{man03-13082009-05cdim}
\Delta(F^{\ominussm ab}) = 1\,,
\quad \
\Delta(F^{abc}) = 2\,,
\quad \
\Delta(F^{\oplussm\ominussm a}) = 2\,,
\quad \
\Delta(F^{\oplussm ab}) = 3\,.
\ee

Poincar\'e algebra symmetries are realized on a space of the field strengths
in a usual way. Realization of dilatation symmetry is given by
\rf{conalggenlis03}, where conformal dimensions of the field strengths are
given in \rf{man03-13082009-05cdim}. All that remains is to find conformal
boost transformations of the field strengths. Making use of the conformal
boost transformations of the gauge fields defined by relations
\rf{man10112009-05}-\rf{man03-13082009-04}, we find the corresponding
conformal boost transformations of the field strengths. Those conformal boost
transformations of the field strengths can be represented as in
\rf{conalggenlis04xxx},\rf{conalggenlis04xxxnew}, where we substitute, in
place of the tensor fields $\Lambda^{a_1\ldots a_n}$, the fields strengths
with the following realization  of the operator $R^a$:
\beq
\label{man03-13082009-11} && \delta_{R^a} F^{\ominussm bc} = 0\,,
\\[5pt]
\label{man03-13082009-12} && \delta_{R^a} F^{bce} = - \eta^{ab}F^{\ominussm
ce} - \eta^{ac}F^{\ominussm eb} - \eta^{ae}F^{\ominussm bc}\,,
\\[5pt]
\label{man03-13082009-13} &&  \delta_{R^a} F^{\oplussm\ominussm b} = -
F^{\ominussm ab}\,,
\\[5pt]
\label{man03-13082009-14} &&   \delta_{R^a} F^{\oplussm bc}= \eta^{ab}
F^{\oplussm\ominussm c}  - \eta^{ac} F^{\oplussm\ominussm b} + F^{abc} -
\partial^a F^{\ominussm bc}\,.
\eeq
From \rf{man03-13082009-11}-\rf{man03-13082009-14}, we see the operator $R^a$
maps the field strength with conformal dimension $\Delta$ into the ones
having conformal dimension less than $\Delta$. In other words, the
realization of the operator $R^a$ given in
\rf{man03-13082009-11}-\rf{man03-13082009-14} can schematically be
represented as
\beq
&  F^\oplussm \stackrel{R}{\longrightarrow} F^{\oplussm\ominussm} \oplus F
\oplus
\partial F^\ominussm\,, &
\\[7pt]
& F \stackrel{R}{\longrightarrow} F^\ominussm\,,\qquad F^{\oplussm\ominussm}
\stackrel{R}{\longrightarrow} F^\ominussm \,,  \qquad F^\ominussm
\stackrel{R}{\longrightarrow} 0\,. &  \eeq

{\it On-shell degrees of freedom and light-cone gauge Lagrangian}. In order
to discuss on-shell D.o.F of the conformal self-dual field we use a
nomenclature of the $so(d-2)$ algebra which is $so(2)$ when $d=4$ . Namely,
we decompose the on-shell D.o.F into irreps of the $so(2)$ algebra. One can
prove that the on-shell D.o.F of the self-dual field are described by two
$so(2)$ algebra self-dual complex-valued vector fields $\phi_{-1}^i$,
$\phi_1^i$ and one complex-valued scalar fields $\phi_0$,

\be \label{11122008-01} \phi_{-1}^i\,, \qquad \phi_{1}^i\,, \qquad
\phi_0\,,\ee
where vector indices of the $so(2)$ algebra take values $i,j=1,2$. The vector
fields satisfy the $so(2)$ self-duality constraint:
\be \label{man11112009-01} \phi_{-1}^i = \tau^{-1} \epsilon^{ij}\phi_{-1}^j
\,,\qquad \phi_1^i = \tau^{-1} \epsilon^{ij}\phi_1^j\,, \ee
where $\epsilon^{ij}$ is the Levi-Civita symbol normalized as
$\epsilon^{12}=1$.

Using light-cone gauge frame, one can make sure that gauge invariant
Lagrangian \rf{080312-23} leads to the following light-cone gauge Lagrangian
for fields \rf{11122008-01}:
\be \label{man03-14082009-01} \LL_{\rm l.c.} = \bar\phi_1^i \Box \phi_{-1}^i
+ \bar\phi_{-1}^i \Box \phi_1^i + \bar\phi_0 \Box \phi_0 - \bar\phi_1^i
\phi_1^i \,. \ee
Details of the derivation of the on-shell D.o.F and light-cone gauge
Lagrangian may be found in Appendix C.

From \rf{11122008-01},\rf{man11112009-01}, we see that number of real-valued
on-shell D.o.F is equal to 6. This result agrees with the one found in
Ref.\cite{Fradkin:1985am}. Note however that we find not only the number of
the on-shell D.o.F but also provide the decomposition of those on-shell D.o.F
into irreps of the $so(2)$algebra.

\newsection{ Ordinary-derivative approach to self-dual field for
arbitrary $d=2\nu$} \label{man02-sec-04}

We now develop ordinary-derivative approach to conformal self-dual field
propagating in flat space-time of arbitrary even dimension $d=2\nu$. To
discuss ordinary-derivative approach to the self-dual field we use the
following field content:
\be \label{05122208-01} \phi_{-1}^{a_1\ldots a_{\nu-1}}\,,\qquad
\phi_{1}^{a_1\ldots a_{\nu-1}}\,,\qquad \phi_0^{a_1\ldots a_{\nu-2}}
\,,\qquad t^{a_1\ldots a_\nu }\,,\ee
where the field  $t^{a_1\ldots a_\nu }$ satisfies the self-duality constraint
\be \label{13122008-01}  t^{a_1\ldots a_\nu } =
\frac{\tau}{\nu!}\epsilon^{a_1\ldots a_\nu b_1\ldots b_\nu } t^{b_1\ldots
b_\nu}\,,\ee
and $\tau$ is defined in \rf{man10112009-01}. We note that:
\\
i)  Fields in \rf{05122208-01} are antisymmetric tensor fields of the Lorentz
algebra $so(d-1,1)$.\\
ii) Fields in \rf{05122208-01} are complex-valued when $d=4k$ and real-valued
when $d=4k+2$.
\\
iii) Conformal dimensions of fields in \rf{05122208-01} are given by

$$
\Delta(\phi_{-1}^{a_1\ldots a_{\nu-1}})=\frac{d-4}{2}\,,\qquad
\Delta(\phi_{1}^{a_1\ldots a_{\nu-1}})=\frac{d}{2}\,,\qquad
\Delta(\phi_0^{a_1\ldots a_{\nu-2}})=\frac{d-2}{2} \,,
$$
\be \label{05122208-02}  \Delta(t^{a_1\ldots a_\nu }) = \frac{d-2}{2}\,.\ee
We note that subscript $k'$ in $\phi_{k'}$ implies that conformal dimension
of the field $\phi_{k'}$ is equal to $\frac{d-2}{2} +k'$.

Ordinary-derivative action we found is given by
\be \label{man03-15082009-01x} S = \int d^dx\, \LL \,,\ee
where Lagrangian takes the form
\beq \label{06122008-03}
\LL &= & -\frac{1}{\nu!} F^{a_1\ldots a_\nu}(\bar\phi_{-1}) F^{a_1\ldots
a_\nu}(\phi_1)
-\frac{1}{\nu!} F^{a_1\ldots a_\nu}(\bar\phi_1) F^{a_1\ldots
a_\nu}(\phi_{-1})
\nonumber\\[5pt]
& - &  \frac{1}{\nu!} \bar{t}^{a_1\ldots a_\nu} F^{a_1\ldots a_\nu}(\phi_1)
-  \frac{1}{\nu!} t^{a_1\ldots a_\nu} F^{a_1\ldots a_\nu}(\bar\phi_1)
\nonumber\\[5pt]
& - & \frac{1}{(\nu-1)!} (\bar\phi_1^{a_1\ldots a_{\nu-1}} + F^{a_1\ldots
a_{\nu-1}}(\bar\phi_0))(\phi_1^{a_1\ldots a_{\nu-1}} + F^{a_1\ldots
a_{\nu-1}}(\phi_0))\,,
\eeq
when $d=4k$, while for $d=4k+2$, Lagrangian is given by
\beq  \label{06122008-04}
\LL &= & -\frac{1}{\nu!} F^{a_1\ldots a_\nu}(\phi_{-1}) F^{a_1\ldots
a_\nu}(\phi_1)
-  \frac{1}{\nu!} t^{a_1\ldots a_\nu} F^{a_1\ldots a_\nu}(\phi_1)
\nonumber\\[5pt]
& - & \frac{1}{2(\nu-1)!} (\phi_1^{a_1\ldots a_{\nu-1}} + F^{a_1\ldots
a_{\nu-1}}(\phi_0))(\phi_1^{a_1\ldots a_{\nu-1}} + F^{a_1\ldots
a_{\nu-1}}(\phi_0))\,,
\eeq
where field strengths are defined as

\be  \label{man11112009-02} F^{a_1\ldots a_n}(\phi)= n
\partial^{[a_1}\phi^{a_2 \ldots a_n]} \ee
and the antisymmetrization of the tensor indices is normalized as $[a_1
\ldots a_n] = \frac{1}{n!} (a_1 \ldots a_n  \pm (n!-1)$ terms).

We note that:
\\
{\bf i}) Two-derivative contributions to Lagrangians
\rf{06122008-03},\rf{06122008-04} take the form of standard second-order
kinetic terms for antisymmetric tensor fields. Besides the two-derivative
contributions, the Lagrangians involve one-derivative contributions and
derivative-independent mass-like contributions.
\\
{\bf ii}) Equations of motion for $t^{a_1\ldots a_\nu}$ and self-duality
constraint for $t^{a_1\ldots a_\nu}$ \rf{13122008-01} imply that {\it
on-shell} the field strength $F^{a_1\ldots a_\nu}(\phi_1)$ satisfies the
self-duality constraint,
\be F^{a_1\ldots a_\nu}(\phi_1) = \frac{\tau}{\nu!}\epsilon^{a_1\ldots a_\nu
b_1\ldots b_\nu } F^{b_1\ldots b_\nu}(\phi_1)\,.
\ee

{\it Gauge transformations}. We now discuss gauge symmetries of Lagrangians
\rf{06122008-03},\rf{06122008-04}. To this end we introduce the following
gauge transformation parameters:
\be \label{06122208-01} \xi_{-2}^{a_1\ldots a_{\nu-2}}\,,\qquad
\xi_0^{a_1\ldots a_{\nu-2}}\,,\qquad \xi_{-1}^{a_1\ldots a_{\nu-3}} \,,\qquad
\lambda^{a_1\ldots a_{\nu-1} }\,.\ee
We note that:
\\
{\bf i})  Gauge transformation parameters \rf{06122208-01} are antisymmetric
tensor
fields of the Lorentz algebra $so(d-1,1)$.\\
{\bf ii}) Gauge transformation parameters \rf{06122208-01} are complex-valued
when $d=4k$ and real-valued when $d=4k+2$.
\\
{\bf iii}) Conformal dimensions of gauge transformation parameters
\rf{06122208-01} are given by
$$
\Delta(\xi_{-1}^{a_1\ldots a_{\nu-2}})=\frac{d-6}{2}\,,\qquad
\Delta(\xi_0^{a_1\ldots a_{\nu-2}})=\frac{d-2}{2}\,,\qquad
\Delta(\xi_{-1}^{a_1\ldots a_{\nu-3}})=\frac{d-4}{2} \,,
$$
\be \label{06122208-02}  \Delta(\lambda^{a_1\ldots a_{\nu-1} }) =
\frac{d-4}{2}\,.\ee

Gauge transformations we found take the form:
\beq
\label{06122208-05} && \delta\phi_1^{a_1\ldots a_{\nu-1}} =  F^{a_1\ldots
a_{\nu-1}}(\xi_0)\,,
\\
\label{06122208-06} && \delta\phi_{-1}^{a_1\ldots a_{\nu-1}} = F^{a_1\ldots
a_{\nu-1}}(\xi_{-2}) - \lambda^{a_1\ldots a_{\nu-1}} \,,
\\
\label{06122208-07} && \delta\phi_0^{a_1\ldots a_{\nu-2}} = F^{a_1\ldots
a_{\nu-2}}(\xi_{-1})- \xi_0^{a_1\ldots a_{\nu-2}} \,,
\\
\label{06122208-08} && \delta t^{a_1\ldots a_\nu} = F^{a_1\ldots
a_\nu}(\lambda) + \frac{\tau}{\nu!} \epsilon^{a_1\ldots a_\nu b_1\ldots
b_\nu}F^{b_1\ldots b_\nu}(\lambda)\,,
\eeq
where strengths for the gauge transformation parameters are defined as in
\rf{man11112009-02}.

From \rf{06122208-06},\rf{06122208-07}, we see that the gauge fields
$\phi_{-1}^{a_1\ldots a_{\nu-1}}$, $\phi_0^{a_1\ldots a_{\nu-2}}$ transform
as Stueckelberg fields, i.e., these fields can be gauged away via
Stueckelberg gauge fixing, $\phi_{-1}^{a_1\ldots a_{\nu-1}}=0$,
$\phi_0^{a_1\ldots a_{\nu-2}}=0$. If we gauge away these fields and exclude
the field $\phi_1^{a_1\ldots a_{\nu-1}}$ via equations of motion, then our
Lagrangians \rf{06122008-03},\rf{06122008-04} reduce to the respective
Lagrangians of the standard approach, \rf{04122008-01},\rf{04122008-01xx}.
Thus, our approach is equivalent to the standard one. Note that
one-derivative contributions to gauge transformations
\rf{06122208-05}-\rf{06122208-08} take the form of the standard gauge
transformations for antisymmetric tensor fields.

{\it Realization of conformal algebra symmetries}.  To complete the
ordinary-derivative description of the conformal self-dual field we should
provide realization of the conformal algebra symmetries on a space of tensor
fields \rf{05122208-01}. Realization of the Poincar\'e algebra symmetries on
the tensor fields is well known. Realization of dilatation symmetry is given
by \rf{conalggenlis03}, where conformal dimensions of fields \rf{05122208-01}
are given in \rf{05122208-02}. General form of conformal boost
transformations of arbitrary tensor fields is given in
\rf{conalggenlis04xxx}. According to \rf{conalggenlis04xxx} we should find
realizations of the operators $K_{\Delta,M}^a$ and $R^a$ on space of fields
\rf{05122208-01}. The realization of the former operator is obtained by
adopting general formula \rf{conalggenlis04xxxnew} for gauge fields
\rf{05122208-01}. The real problem is getting a realization of the operator
$R^a$. The realization of the operator $R^a$ on space of gauge fields
\rf{05122208-01} we found is given by
\beq \label{10122208-08}
\delta_{R^a} \phi_1^{a_1\ldots a_{\nu-1}} & = &- t^{a a_1\ldots a_{\nu-1}}
-(\nu-1)\eta^{a[a_1}\phi_0^{a_2\ldots a_{\nu-1}]}
\nonumber\\[5pt]
& - & \partial^a \phi_{-1}^{a_1\ldots a_{\nu-1}} -
\frac{\tau}{\nu!}\epsilon^{aa_1\ldots a_{\nu-1}b_1\ldots b_\nu}F^{b_1\ldots
b_\nu}(\phi_{-1})\,,
\\[7pt]
\delta_{R^a} \phi_{-1}^{a_1\ldots a_{\nu-1}} & = & 0\,,
\\[7pt]
\delta_{R^a} \phi_0^{a_1\ldots a_{\nu-2}} & = & \phi_{-1}^{aa_1\ldots
a_{\nu-2}}\,,
\\[7pt]
\label{10122208-14}
\delta_{R^a} t^{a_1\ldots a_\nu} & = & \nu \eta^{a[a_1}\phi_{-1}^{a_2\ldots
a_\nu]}
+ \frac{\tau}{(\nu-1)!}\epsilon^{a_1\ldots a_\nu a b_1\ldots b_{\nu-1}}
\phi_{-1}^{b_1\ldots b_{\nu-1}}\,.
\eeq
Making use of these relations and general formula \rf{conalggenlis04xxx}
gives the conformal boost transformations of gauge fields \rf{05122208-01}.

{\it On-shell degrees of freedom and light-cone gauge Lagrangian}. We now
discuss on-shell D.o.F of the self-dual field. As before, for this purpose it
is convenient to use fields transforming in irreps of the $so(d-2)$ algebra.
Using the method in Appendix C, one can prove that on-shell D.o.F are
described by the following antisymmetric tensor fields of the $so(d-2)$
algebra:
\be \label{11122008-03}
\phi_{-1}^{i_1\ldots i_{\nu-1}}\,,\qquad \phi_1^{i_1\ldots
i_{\nu-1}}\,,\qquad
\phi_0^{i_1\ldots i_{\nu-2}}\,,  \ee
where vector indices of the $so(d-2)$ algebra take values $i,j=1,2,\ldots
d-2$. Fields in \rf{11122008-03} are complex-valued when $d=4k$ and
real-valued when $d=4k+2$. The fields $\phi_{-1}^{i_1\ldots i_{\nu-1}}$,
$\phi_1^{i_1\ldots i_{\nu-1}}$ satisfy the $so(d-2)$ self-duality constraint,
\beq
&& \phi_{-1}^{i_1\ldots i_{\nu-1}} =
\frac{\tau^{-1}}{(\nu-1)!}\epsilon^{i_1\ldots i_{\nu-1} j_1 \ldots j_{\nu-1}}
\phi_{-1}^{j_1\ldots j_{\nu-1}}\,,
\\[3pt]
&& \phi_1^{i_1\ldots i_{\nu-1}} =
\frac{\tau^{-1}}{(\nu-1)!}\epsilon^{i_1\ldots i_{\nu-1} j_1 \ldots j_{\nu-1}}
\phi_1^{j_1\ldots j_{\nu-1}}\,, \eeq
where $\epsilon^{i_1\ldots i_{\nu-1} j_1 \ldots j_{\nu-1}}$ is the
Levi-Civita symbol normalized as $\epsilon^{12\ldots d-2}=1$.

Total number of real-valued on-shell D.o.F given in \rf{11122008-03} is equal
to
\be \label{DOFspi01} \nbf = \frac{ h (5\nu-7)(2\nu-4)!}{(\nu-1)!(\nu-2)!}\,,
\qquad
h =  \left\{\begin{array}{l}
2 \ \ \hbox{ for } \ \ d = 4k;
\\[7pt]
1 \ \ \hbox{ for } \ \ d = 4k+2\,.
\end{array}\right.
\ee
Namely, we note that $\nbf$ is a sum of $\nbf(\phi_{\pm 1}^{i_1\ldots
i_{\nu-1}})$ and $\nbf(\phi_0^{i_1\ldots i_{\nu-2}})$ which are the
respective numbers of the real-valued independent tensorial components of the
fields $\phi_{\pm 1}^{i_1\ldots i_{\nu-1}}$ and $\phi_0^{i_1\ldots
i_{\nu-2}}$,
\beq
&& \nbf = \nbf (\phi_1^{i_1\ldots i_{\nu-1}}) + \nbf (\phi_{-1}^{i_1\ldots
i_{\nu-1}}) + \nbf(\phi_0^{i_1\ldots i_{\nu-2}}) \,,
\\[3pt]
&& \nbf(\phi_{\pm 1}^{i_1\ldots i_{\nu-1}}) =
\frac{h(2\nu-2)!}{2((\nu-1)!)^2}\,,
\qquad  \nbf(\phi_0^{i_1\ldots i_{\nu-2}}) =
\frac{h(2\nu-4)!}{((\nu-2)!)^2}\,.
\eeq

Using light-cone gauge frame, one can make sure that, for $d=4k$, gauge
invariant Lagrangian \rf{06122008-03} leads to the following light-cone gauge
Lagrangian for fields \rf{11122008-03}:
\beq \label{12122008-01}
\LL_{\rm l.c.} & = & \frac{1}{(\nu-1)!} \bar\phi_1^{i_1\ldots i_{\nu-1}}\Box
\phi_{-1}^{i_1\ldots i_{\nu-1}} + \frac{1}{(\nu-1)!} \bar\phi_{-1}^{i_1\ldots
i_{\nu-1}}\Box \phi_1^{i_1\ldots i_{\nu-1}}
\nonumber\\[5pt]
& + &   \frac{1}{(\nu-2)!}  \bar\phi_0^{i_1\ldots i_{\nu-2}} \Box
\phi_0^{i_1\ldots i_{\nu-2}} - \frac{1}{(\nu-1)!} \bar\phi_1^{i_1\ldots
i_{\nu-1}}\phi_1^{i_1\ldots i_{\nu-1}}\,,
\eeq
while, for $d=4k+2$, gauge invariant Lagrangian \rf{06122008-04} leads to the
following light-cone gauge Lagrangian for fields \rf{11122008-03}:
\beq  \label{12122008-02}
\LL_{\rm l.c.} & = & \frac{1}{(\nu-1)!} \phi_1^{i_1\ldots i_{\nu-1}}\Box
\phi_{-1}^{i_1\ldots i_{\nu-1}} +  \frac{1}{2(\nu-2)!}  \phi_0^{i_1\ldots
i_{\nu-2}} \Box \phi_0^{i_1\ldots i_{\nu-2}}
\nonumber\\[5pt]
& - & \frac{1}{2(\nu-1)!}\phi_1^{i_1\ldots i_{\nu-1}}\phi_1^{i_1\ldots
i_{\nu-1}}\,.
\eeq

\newsection{ Oscillator form of Lagrangian}\label{man02-sec-05}

In the preceding sections, we have presented our results for the self-dual
fields by using the representation of field content in terms of the tensor
fields. However use of such representation is not convenient in many
applications. In this section, we represent our results by using the
representation of field content in terms of generating functions constructed
out of the tensor fields and
the appropriate oscillators.%
\footnote{ Note that in this paper we use oscillators just to handle the many
indices appearing for tensor fields. In a proper way, the oscillators arise
in the framework of world-line approach to higher-spin fields (see e.g.
Refs.\cite{Engquist:2005yt}-\cite{Bastianelli:2008nm}.)}
This is to say that in order to obtain the Lagrangian description in an
easy--to--use form, we introduce creation operators $\alpha^a$, $\zeta$,
$\upsilon^\oplussm$, $\upsilon^\ominussm$ and the respective annihilation
operators $\bar\alpha^a$, $\bar\zeta$, $\bar\upsilon^\ominussm$,
$\bar\upsilon^\oplussm$ and collect tensor fields \rf{05122208-01} in
ket-vector $\Phik$ defined by
\be \label{10122208-01} \Phik = \left( \begin{array}{l}
\phik
\\[5pt]
\tk
\end{array}\right)\,,
\ee

\beq \label{02112009man-01}
&& \phik =  \upsilon^\oplussm |\phi_1 \rangle + \upsilon^\ominussm |\phi_{-1}
\rangle + \zeta |\phi_0\rangle \,,
\\[7pt]
\label{02112009man-02} && |\phi_{k'}\rangle \equiv
\frac{1}{(\nu-1)!}\alpha^{a_1} \ldots \alpha^{a_{\nu-1}} \phi_{k'}^{a_1\ldots
a_{\nu-1}} |0\rangle\,,\qquad k'=-1,1\,,
\\[7pt]
\label{02112009man-03} && |\phi_0\rangle \equiv \frac{1}{(\nu-2)!}
\alpha^{a_1} \ldots \alpha^{a_{\nu-2}} \phi_{0}^{a_1\ldots a_{\nu-2}}
|0\rangle\,,
\\[7pt]
\label{02112009man-04} && \tk \equiv \frac{1}{\nu!} \alpha^{a_1} \ldots
\alpha^{a_\nu} t^{a_1\ldots a_\nu} |0\rangle\,.
\eeq
In the literature, ket-vectors \rf{10122208-01}-\rf{02112009man-04} are
sometimes referred to as generating functions. The ket-vectors $\phik$, $\tk$
satisfy the obvious algebraic constraints
\beq
&& (N_\alpha + N_\zeta)\phik =  (\nu -1) \phik\,,
\\[5pt]
&& (N_\zeta + N_\upsilon)\phik = \phik\,,
\\[5pt]
&& N_\alpha \tk =  \nu \tk\,,
\\[5pt]
&& N_\zeta \tk = 0\,,\qquad
N_\upsilon\tk = 0\,,
\eeq
where we use the notation given in \rf{10122208-02}-\rf{10122208-07}. We note
that these algebraic constraints tell us about number of the oscillators
$\alpha^a$, $\zeta$, $\upsilon^\oplussm$, $\upsilon^\ominussm$ appearing in
the ket-vectors $\phik$ and $\tk$. In terms of the ket-vector $\tk$,
self-duality constraint \rf{13122008-01} takes the form
\be \label{man10112009-02} \tk =\epsilon \tk\,, \ee
where we use the notation for $\epsilon$-symbol
\be \label{080313-41}  \epsilon \equiv \frac{\tau}{(\nu!)^2}
\epsilon^{a_1\ldots a_\nu b_1 \ldots b_\nu}\alpha^{a_1}\ldots \alpha^{a_\nu}
\bar\alpha^{b_\nu}\ldots \bar\alpha^{b_1}\,.
\ee
Useful relations for $\epsilon$-symbol \rf{080313-41} and various related
$\epsilon$-symbols may be found in Appendix D.

In terms of the ket-vector $\Phik$, Lagrangians
\rf{06122008-03},\rf{06122008-04} can be reexpressed as
\be \label{man03-15082009-01} \LL = \frac{h}{2} \Phibr E \Phik \,, \ee
where the normalization factor $h$ is given in \rf{DOFspi01}, while operator
$E$ is defined by the relations
\be \label{man03-15082009-02} E = \left( \begin{array}{ll}
E_{\phi\phi} & E_{\phi t}
\\[5pt]
E_{t\phi} &  0
\end{array}\right)\,,
\ee
\beq
\label{man03-15082009-03} && E_{\phi\phi} = E_{\phi\phi\smtwo} +
E_{\phi\phi\smone} + E_{\phi\phi\smzero} \,,
\\[5pt]
\label{man03-15082009-04} && E_{\phi t} = \upsilon^\ominussm \albpar\,,
\qquad
E_{t \phi} = -\bar\upsilon^\ominussm \alpar\,,
\eeq
\beq
\label{man03-15082009-06} && E_{\phi\phi\smtwo} = \Box - \alpar\albpar\,,
\\[5pt]
\label{man03-15082009-07} && E_{\phi\phi\smone} =  e_1 \albpar + \alpar \eb_1
\,,
\\[5pt]
\label{man03-15082009-08} && E_{\phi\phi\smzero} = m_1\,,
\\[7pt]
\label{man03-15082009-09} && e_1 = \zeta  \bar\upsilon^\ominussm\,,
\qquad
\eb_1 = -\upsilon^\ominussm   \bar\zeta\,,
\qquad
\label{man03-15082009-11} m_1 = \upsilon^\ominussm  \bar\upsilon^\ominussm
(N_\zeta-1)\,.
\eeq

Alternatively, Lagrangian \rf{man03-15082009-01} can be represented in terms
of ket-vector of gauge fields \rf{02112009man-01}-\rf{02112009man-04}. This
is to say that the Lagrangian takes the form (up to a total derivative)
\beq \label{man-29112009-01}
- \LL & = &  \langle F(\phi_{-1})|F(\phi_1)\rangle + \langle
F(\phi_1)|F(\phi_{-1})\rangle
+ \langle F(\phi_1)\tk  + \tbr F(\phi_1)\rangle
\nonumber\\[5pt]
& + & \Bigl( \langle\phi_1| + \langle F(\phi_0)| \Bigr)\Bigl(|\phi_1\rangle +
|F(\phi_0)\rangle\Bigr)\,,
\eeq
when $d=4k$, and
\beq \label{man-29112009-02}
- \LL & = &  \langle F(\phi_{-1})|F(\phi_1)\rangle +  \langle F(\phi_1)\tk
\nonumber\\[5pt]
& + & \half \Bigl( \langle\phi_1| + \langle F(\phi_0)|
\Bigr)\Bigl(|\phi_1\rangle + |F(\phi_0)\rangle\Bigr)\,,
\eeq
when $d=4k+2$. In \rf{man-29112009-01}, \rf{man-29112009-02} and below,
ket-vector of field strength $|F(\phi)\rangle$ is defined as
\be \label{man-12112000-01} |F(\phi)\rangle =\alpar \phik\,. \ee

{\it Gauge transformations}. Gauge transformations can also be cast into the
generating form. To this end we use, as before, the oscillators $\alpha^a$,
$\zeta$, $\upsilon^\oplussm$, $\upsilon^\ominussm$ and collect gauge
transformation parameters \rf{06122208-01} into ket-vector $\Xik$ defined by
\be \Xik = \left( \begin{array}{l}
\xik
\\[5pt]
\lambdak
\end{array}\right)\,,
\ee
where ket-vectors $\xik$ and $\lambdak$ are defined as

\beq
&& \xik = \upsilon^\oplussm  |\xi_0 \rangle + \upsilon^\ominussm |\xi_{-2}
\rangle - \zeta |\xi_{-1}\rangle\,,
\\[10pt]
&& |\xi_{k'-1}\rangle \equiv \frac{1}{(\nu-2)!} \alpha^{a_1} \ldots
\alpha^{a_{\nu-2}} \xi_{k'-1}^{a_1\ldots a_{\nu-2}} |0\rangle\,,\qquad
k'=-1,1\,,
\\[15pt]
&& |\xi_{-1}\rangle \equiv \frac{1}{(\nu-3)!} \alpha^{a_1} \ldots
\alpha^{a_{\nu-3}} \xi_{-1}^{a_1\ldots a_{\nu-3}} |0\rangle\,,
\\[15pt]
&& \lambdak \equiv \frac{1}{(\nu-1)!} \alpha^{a_1} \ldots \alpha^{a_{\nu-1}}
\lambda^{a_1\ldots a_{\nu-1}} |0\rangle\,.
\eeq
The ket-vectors $\xik$, $\lambdak$ satisfy the obvious algebraic constraints
\beq \label{coco-01}
&& (N_\alpha + N_\zeta)\xik =  (\nu -2) \xik\,,
\\[5pt]
\label{coco-02} && (N_\zeta + N_\upsilon)\xik = \xik\,,
\\[5pt]
\label{coco-03} && N_\alpha \lambdak =  (\nu-1) \lambdak\,,
\\[5pt]
\label{coco-04} && N_\zeta\lambdak = 0\,, \qquad N_\upsilon\lambdak = 0\,.
\eeq
As before, these constraints tell us about number of the oscillators $\alpha^a$,
$\zeta$, $\upsilon^\oplussm$, $\upsilon^\ominussm$ appearing in the
ket-vectors $\xik$ and $\lambdak$.

Now, gauge transformations \rf{06122208-05}-\rf{06122208-08} can entirely be
represented in terms of the ket-vectors $\Phik$ and $\Xik$,
\be \label{02112009man-09}\delta \Phik    =   G \Xik  \,,
\ee
where the operator $G$ is given by
\be G = \left( \begin{array}{cc}
\alpar - \zeta \bar\upsilon^\ominussm \qquad &  - \upsilon^\ominussm
\\[7pt]
0 \qquad &  (1+\epsilon)\alpar
\end{array}\right)\,,
\ee
and the $\epsilon$-symbol is defined in \rf{080313-41}.

Alternatively, gauge transformation \rf{02112009man-09} can be represented in
terms of ket-vector of gauge fields \rf{02112009man-01}-\rf{02112009man-04}.
This is to say that gauge transformation \rf{02112009man-09} amounts to the
following gauge transformations
\beq \label{man-15112009-01}
&& \delta |\phi_1\rangle  = |F(\xi_0)\rangle\,,
\\[5pt]
\label{man-15112009-02} && \delta |\phi_{-1}\rangle  =  |F(\xi_{-2})\rangle -
|\lambda\rangle\,,
\\[5pt]
\label{man-15112009-03} && \delta |\phi_0\rangle  =  |F(\xi_{-1})\rangle -
|\xi_0\rangle\,,
\\[5pt]
\label{man-15112009-04} && \delta \tk = (1+\epsilon) |F(\lambda)\rangle\,,
\eeq
where the field strengths for ket-vectors of the gauge transformation
parameters are defined as in \rf{man-12112000-01}.

{\it Oscillator realization of conformal algebra symmetries on gauge fields}.
To complete the oscillator description of the conformal self-dual field we
provide a realization of the conformal algebra symmetries on space of the
ket-vector $\Phik$. Realization of the Poincar\'e algebra symmetries and
dilatation symmetry is given by \rf{conalggenlis01}-\rf{conalggenlis03},
where the operators $M^{ab}$ and $\Delta$ take the form
\beq
\label{man-12112009-04} &&  \label{080313-44xx} M^{ab} =\alpha^a\bar\alpha^b
- \alpha^b\bar\alpha^a\,,
\\[3pt]
\label{080313-44x}
&& \Delta  =  \frac{d-2}{2} +  \Delta'\,,\qquad \Delta' \equiv
N_{\upsilon^\oplussm} - N_{\upsilon^\ominussm}\,.
\eeq
Conformal boost transformations of $\Phik$ are given in \rf{conalggenlis04}.
According to \rf{conalggenlis04} we should find the operators
$K_{\Delta,M}^a$ and $R^a$. The former operator is given in
\rf{conalggenlis05}. Realization of the operator $R^a$ on space of $\Phik$
can be read from \rf{10122208-08}-\rf{10122208-14}. Namely, in terms of the
ket-vector $\Phik$, transformations of the gauge fields given in
\rf{10122208-08}-\rf{10122208-14} can be represented as
\be \label{02112009man-17} \delta_{R^a}\Phik = R^a \Phik \,,\ee
where the realization of the operator $R^a$ on $\Phik$ takes the form
\be \label{02112009man-16} R^a = \left( \begin{array}{cc}
R_{\phi\phi}^a  &  R_{\phi t}^a
\\[7pt]
R_{t\phi}^a  &  0
\end{array}\right)\,,
\ee

\beq
&& R_{\phi\phi }^a = r_{0,1}\bar\alpha^a  + \alpha^a \rb_{0,1}
+ r_{1,1}(\eta^{ab} + \epsilon_0^{ab})\partial^b\,,
\\[7pt]
&& R_{\phi t}^a = r_{0,4}\bar\alpha^a\,,
\\[7pt]
&& R_{ t\phi}^a = r_{0,5}(1+\epsilon)\alpha^a\,,
\\[7pt]
&& \qquad r_{0,1} = \zeta  \bar\upsilon^\oplussm\,,
\qquad \rb_{0,1} = - \upsilon^\oplussm  \bar\zeta\,,
\\
&& \qquad r_{1,1} = - \upsilon^\oplussm \bar\upsilon^\oplussm\,,
\\
&& \qquad r_{0,4} = - \upsilon^\oplussm \,,
\qquad r_{0,5} =  \bar\upsilon^\oplussm\,,
\eeq
and $\epsilon_0^{ab}$-symbol is defined in \rf{080313-44}.

Alternatively, conformal boost transformations \rf{02112009man-17} can be
represented in terms of ket-vectors \rf{02112009man-01}-\rf{02112009man-04}.
To this end we note that realization of the operator $K_{\Delta,M}^a$ on
space of ket-vectors \rf{02112009man-01}-\rf{02112009man-04} is given by
\rf{conalggenlis05}, where the operators $M^{ab}$ and $\Delta$ take the same
form as in \rf{man-12112009-04},\rf{080313-44x}. Note that \rf{080313-44x}
implies the following conformal dimensions of the respective ket-vectors
\beq \label{02112009man-21}
&& \Delta(|\phi_{k'}\rangle) = \frac{d-2}{2}+k' \,, \qquad k'=0,\pm1\,,
\\
\label{02112009man-22} &&  \Delta(\tk) = \frac{d-2}{2}\,.\eeq
We now make sure that the realization of the operator $R^a$ given in
\rf{02112009man-17}  can be represented in terms of ket-vectors
\rf{02112009man-01}-\rf{02112009man-04} as
\beq \label{02112009man-10}
&& \delta_{R^a}|\phi_1\rangle = -\bar\alpha^a\tk -\alpha^a |\phi_0\rangle -
(\eta^{ab}+\epsilon_0^{ab})\partial^b |\phi_{-1}\rangle\,,
\\[5pt]
\label{02112009man-11} && \delta_{R^a}|\phi_{-1}\rangle =0\,,
\\[5pt]
\label{02112009man-12} && \delta_{R^a} |\phi_0\rangle = \bar\alpha^a
|\phi_{-1}\rangle\,,
\\[5pt]
\label{02112009man-13} && \delta_{R^a} |t\rangle = (1+\epsilon)\alpha^a
|\phi_{-1}\rangle\,.
\eeq
Making use of these relations gives the conformal boost transformations of
ket-vectors \rf{02112009man-01}-\rf{02112009man-04}.

We now remind that the realization of the conformal symmetries on a space of
the ket-vectors of gauge fields \rf{02112009man-10}-\rf{02112009man-13} is
defined up to gauge transformations
\rf{man-15112009-01}-\rf{man-15112009-04}. As we have demonstrated in Sec. 3,
the conformal symmetries can also be realized on a space of the field
strengths. We now discuss the oscillator form of the field strengths for
gauge fields \rf{02112009man-01}-\rf{02112009man-04} and the corresponding
realization of conformal symmetries on a space of the field strengths.

{\it Oscillator realization of conformal algebra symmetries on field
strengths}. We introduce the following ket-vectors of field strengths
constructed out of ket-vectors of gauge fields
\rf{02112009man-01}-\rf{02112009man-04}:
\beq \label{02112009man-05}
&& | F^{\ominussm}\rangle  = \tk + (1+\epsilon) |F(\phi_{-1})\rangle\,,
\\[5pt]
\label{02112009man-06} && |F\rangle  = \alpar |F^\ominussm\rangle \,,
\\[5pt]
\label{02112009man-07} && | F^{\oplussm\ominussm}\rangle  = |\phi_1\rangle  +
|F(\phi_0)\rangle \,,
\\[5pt]
\label{02112009man-08} && | F^{\oplus}\rangle  = |F(\phi_1)\rangle \,,
\eeq
where $|F(\phi_{k'})\rangle$,  $k'=0,\pm1$ are defined as in
\rf{man-12112000-01}. One can make sure that field strengths
\rf{02112009man-05}-\rf{02112009man-08} are invariant under gauge
transformations \rf{man-15112009-01}-\rf{man-15112009-04}.

Conformal dimensions of the field strengths can be read from
\rf{02112009man-21},\rf{02112009man-22} and
\rf{02112009man-05}-\rf{02112009man-08},
\be \label{02112009man-15}
\Delta(| F^{\ominussm}\rangle) = \frac{d-2}{2}\,,
\quad \
\Delta(|F\rangle) = \frac{d}{2}\,,
\quad \
\Delta(| F^{\oplussm\ominussm}\rangle) = \frac{d}{2}\,,
\quad \
\Delta(| F^{\oplus}\rangle) = \frac{d+2}{2}\,.
\ee
Realization of the Poincar\'e algebra symmetries and dilatation symmetry on
space of the ket-vectors of field strengths is given by
\rf{conalggenlis01}-\rf{conalggenlis03}, where the operator $M^{ab}$ takes
the form as in \rf{man-12112009-04}, while conformal dimensions are given in
\rf{02112009man-15}.

Making use of transformations of the ket-vectors of gauge fields given in
\rf{02112009man-10}-\rf{02112009man-13}, we find the corresponding conformal
boost transformations of the ket-vectors of
field strengths,%
\beq
&& \delta_{K^a} |F^{\ominussm }\rangle = K_{\Delta,M}^a
|F^{\ominussm}\rangle\,,
\\[5pt]
&& \delta_{K^a} |F\rangle = K_{\Delta,M}^a |F\rangle  - \alpha^a |
F^\ominussm\rangle\,,
\\[5pt]
&& \delta_{K^a}|F^{\oplussm\ominussm}\rangle = K_{\Delta,M}^a
|F^{\oplussm\ominussm }\rangle -\bar\alpha^a |F^\ominussm\rangle \,,
\\[5pt]
&& \delta_{K^a} | F^{\oplus}\rangle = K_{\Delta,M}^a |F^{\oplus}\rangle
+\alpha^a |F^{\oplussm\ominussm}\rangle + \bar\alpha^a |F\rangle - \partial^a
|F^\ominussm\rangle\,,
\eeq
where the operator $K_{\Delta,M}^a$ is given in \rf{conalggenlis05}, while
the conformal dimensions are defined in \rf{02112009man-15}. Comparing these
formulas with general relation \rf{conalggenlis04}, we find the realization
of the operator $R^a$ on a space of the ket-vectors of the field strengths,
\beq
&& \delta_{R^a} |F^{\ominussm }\rangle = 0\,,
\\[5pt]
&& \delta_{R^a} |F\rangle =  - \alpha^a | F^\ominussm\rangle\,,
\\[5pt]
&& \delta_{R^a}|F^{\oplussm\ominussm}\rangle = - \bar\alpha^a
|F^\ominussm\rangle \,,
\\[5pt]
&& \delta_{R^a} | F^{\oplus}\rangle = \alpha^a |F^{\oplussm\ominussm}\rangle
+ \bar\alpha^a |F\rangle - \partial^a |F^\ominussm\rangle\,.
\eeq

{\it Oscillator form of the light-cone gauge Lagrangian}. To discuss
oscillator form of light-cone gauge Lagrangian we collect fields
\rf{11122008-03} into the following ket-vectors:
\beq \label{02112009man-18}
&& |\phi_{\rm l.c.}\rangle =  \upsilon^\oplussm |\phi_1 \rangle_{\rm l.c} +
\upsilon^\ominussm |\phi_{-1} \rangle_{\rm l.c} + \zeta |\phi_0\rangle_{\rm
l.c} \,,
\\[5pt]
\label{02112009man-19} && |\phi_{k'}\rangle_{\rm l.c} \equiv
\frac{1}{(\nu-1)!}\alpha^{i_1} \ldots \alpha^{i_{\nu-1}} \phi_{k'}^{i_1\ldots
i_{\nu-1}} |0\rangle\,,\qquad k'=-1,1\,,
\\[5pt]
\label{02112009man-20} && |\phi_0\rangle_{\rm l.c} \equiv \frac{1}{(\nu-2)!}
\alpha^{i_1} \ldots \alpha^{i_{\nu-2}} \phi_{0}^{i_1\ldots i_{\nu-2}}
|0\rangle\,.
\eeq
In terms of the ket-vector $|\phi_{\rm l.c.}\rangle$ \rf{02112009man-18},
light-cone gauge Lagrangians \rf{12122008-01},\rf{12122008-02} can concisely
be represented as
\beq
&&  \LL_{\rm l.c.} = \frac{h}{2} \langle \phi_{\rm l.c.}|(\Box - \MM^2
)|\phi_{\rm l.c.}\rangle\,,
\\[3pt]
&& \hspace{1.5cm} \MM^2 \equiv \upsilon^\ominussm \bar\upsilon^\ominussm \,,
\eeq
where the normalization factor $h$ is given in \rf{DOFspi01}.

\newsection{Conclusions}\label{conl-sec-01}

In this paper, we applied the ordinary-derivative approach, developed in
Ref.\cite{Metsaev:2007fq}, to the study of conformal self-dual fields in flat
space of even dimension. The results presented here should have a number of
interesting applications and generalizations, some of which are:

i) Results in this paper and the ones in Ref.\cite{Metsaev:2007fq} provide
the complete ordinary-derivative description of  all fields that appear in
the graviton supermultiplets of conformal supergravity theories. It would be
interesting to apply these results to the study of supersymmetric conformal
field theories \cite{Bergshoeff:1980is}-\cite{Bergshoeff:1986wc} in the
framework of ordinary-derivative approach. The first step in this direction
would be understanding of how the supersymmetries are realized in the
framework of our approach.

ii) Our approach to conformal theories (see
Refs.\cite{Metsaev:2007fq,Metsaev:2008fs}) is based on the new realization of
conformal gauge symmetries via Stueckelberg fields. In our approach, use of
the Stueckelberg fields is very similar to the one in gauge invariant
formulation of  massive fields. Stueckelberg fields provide interesting
possibilities for the study of interacting massive gauge fields (see e.g.
Refs.\cite{Zinoviev:2006im,Metsaev:2006ui}). So we think that application of
our approach to the interacting conformal self-dual fields may lead to new
interesting development.

iii) BRST approach is one of powerful approaches to the analysis of various
aspects of relativistic dynamics (see e.g.
Refs.\cite{Siegel:1999ew,Fotopoulos:2008ka}). This approach is conveniently
adapted for the ordinary-derivative formulation. Recent application of BRST
approach to the study of totally antisymmetric fields may be found in
\cite{Buchbinder:2008kw}. We think that extension of this approach to the
case of conformal self-dual fields should be relatively straightforward.

iv) Self-dual fields studied in this paper are the particular case of
mixed-symmetry fields. In the last years, there were interesting developments
in the studying mixed-symmetry fields
\cite{Zinoviev:2002ye}-\cite{Moshin:2007jt} that are invariant with respect
to Poincar\'e algebra symmetries. It would be interesting to apply methods
developed in Refs.\cite{Zinoviev:2002ye}-\cite{Moshin:2007jt} to the studying
conformal self-dual mixed-symmetry fields%
\footnote{ Unfolded form of equations of motion for conformal mixed-symmetry
fields is studied in Ref.\cite{Shaynkman:2004vu}. Higher-derivative
Lagrangian formulation of the mixed-symmetry conformal fields was recently
developed in Ref.\cite{Vasiliev:2009ck}.}.
There are other various interesting approaches in the literature which could
be used to discuss the ordinary-derivative formulation of conformal self-dual
fields. This is to say that various recently developed interesting
formulations in terms of unconstrained fields in flat space may be found in
Refs.\cite{Buchbinder:2008ss,Campoleoni:2008jq}.

\bigskip

{\bf Acknowledgments}. This work was supported by the RFBR Grant
No.08-02-00963, RFBR Grant for Leading Scientific Schools, Grant No.
1615.2008.2, by the Dynasty Foundation and by the Alexander von Humboldt
Foundation Grant PHYS0167.

\setcounter{section}{0}\setcounter{subsection}{0}
\appendix{ Derivation of ordinary-derivative gauge invariant Lagrangian
 }

Because the methods for finding the ordinary-derivative Lagrangian for
arbitrary $d\geq 4 $ are quite similar we present details of the derivation
for the case of $d=4$. To derive ordinary-derivative gauge invariant
Lagrangian \rf{080312-23} we use Lagrangian of the standard formulation given
in \rf{standlag01}. First, in place of the field $T^{ab}$, we introduce the
fields $t^{ab}$ and $\phi_{-1}^a$ by the relation
\be \label{080312-28} T^{ab} = t^{ab} + F^{ab}(\phi_{-1}) + \frac{\tau
}{2}\epsilon^{abce} F^{ce}(\phi_{-1})\,, \ee
where $t^{ab}$ satisfies self-duality constraint \rf{08122208-02}. Plugging
\rf{080312-28} in \rf{standlag01} we obtain
\beq
\label{080312-30} \LL_{\rm st} &= & -\half F^{ab}(\bar\phi_{-1}) \Box
F^{ab}(\phi_{-1})
\nonumber\\[5pt]
&- & \half \bar{t}^{ab} \Box F^{ab}(\phi_{-1})-\half F^{ab}(\bar\phi_{-1})
\Box t^{ab} +\partial^a \bar{t}^{ac} \partial^b t^{bc}\,,
\eeq
where the field strength $F^{ab}(\phi_{-1})$ is defined as in
\rf{04122208-02}. We note that representation for $T^{ab}$ given in
\rf{080312-28} implies that $T^{ab}$ is invariant under the gauge
transformations
\beq
\label{080312-35} && \delta\phi_{-1}^a = \partial^a \xi_{-2} - \lambda^a \,,
\\
\label{080312-36} && \delta t^{ab} = \partial^a \lambda^b -\partial^b
\lambda^a + \tau \epsilon^{abce} \partial^c \lambda^e\,.
\eeq
This implies that Lagrangian \rf{080312-30} is also invariant under gauge
transformations \rf{080312-35}, \rf{080312-36}.

Second, we introduce new fields $\phi_1^a$ and $\phi_0$ by using the
following Lagrangian in place of \rf{080312-30}
\be
\label{080312-32} \LL = \LL_{\rm st}  - \bar{X}^a X^a\,,
\ee
where we use the notation

\be \label{080312-33} X^a \equiv \phi_1^a + \partial^a \phi_0
-\partial^bF^{ba}(\phi_{-1}) -
\partial^b t^{ba}\,,   \ee
and $\bar{X}^a$ is complex conjugate of $X^a$. It is clear that, on-shell,
Lagrangians $\LL_{\rm st}$ and $\LL$ describe the same field D.o.F.   Using
the formula (up to total derivative)
\be \label{080312-34}  \partial^a t^{ac} \partial^b F^{bc} = - \half
t^{ab}\Box F^{ab}\,,\ee
it is easy to see that Lagrangian \rf{080312-32} gives ordinary-derivative
Lagrangian \rf{080312-23}.

We now consider gauge symmetries of Lagrangian \rf{080312-32}. Because the
contribution to $X^a$ given by (see \rf{080312-33})
\be - \partial^b F^{ba}(\phi_{-1}) - \partial^b t^{ba} \ee
is invariant under gauge transformations \rf{080312-35},\rf{080312-36}, the
$X^a$ is also invariant under these gauge transformations, i.e., Lagrangian
\rf{080312-32} is invariant under gauge transformations
\rf{080312-35},\rf{080312-36}. Besides this, we note that $X^a$ is invariant
under the additional gauge transformations
\be \label{13122008-02}
\delta\phi_1^a = \partial^a \xi_0 \,,
\qquad
\delta\phi_0 = - \xi_0 \,.
\ee
Altogether, gauge transformations
\rf{080312-35},\rf{080312-36},\rf{13122008-02} amount to the ones given in
\rf{080312-24}-\rf{080312-27}.

\appendix{  Derivation of operator $R^a$  }

In this Appendix, we outline the derivation of the operator $R^a$
\rf{02112009man-16}. Use of the oscillator formulation turns out to be
convenient for this purpose. The operator $R^a$ is then determined by
requiring action of self-dual field \rf{man03-15082009-01x} to be invariant
under the conformal boost transformations. In order to analyze restrictions
imposed on the operator $R^a$ by the conformal boost symmetries we need to
know explicit form of the restrictions imposed on operator $E$
\rf{man03-15082009-02} by the Lorentz and dilatation symmetries. Requiring
action \rf{man03-15082009-01x} with Lagrangian \rf{man03-15082009-01} to be
invariant under the Lorentz and dilatation symmetries $\delta_{J^{ab}} S =
0$, $\delta_D S = 0$, amounts to the following respective equations for the
operator $E$:
\be \label{app3-15082009-01x} [E,J^{ab}]=0\,,\qquad [E,D] = 2 E\,,
\ee
where the operators $J^{ab}$ and $D$ are given in \rf{conalggenlis02} and
\rf{conalggenlis03} respectively.

Variation of Lagrangian \rf{man03-15082009-01} under the conformal boost
transformation can be presented as (up to total derivative)
\beq \label{app3-15082009-01}
&& \delta_{K^a} \LL = \frac{h}{2} \Phibr \delta_{K^a} E\Phik \,,
\\[5pt]
\label{app3-15082009-02} && \delta_{K^a} E \equiv  K^{a\dagger} E + E K^a\,.
\eeq
Using \rf{app3-15082009-01x} and $K^a$ given in \rf{conalggenlis04}, we make
sure that $\delta_{K^a} E$ \rf{app3-15082009-02} can be represented as
\beq
\label{delKAEbas01} \delta_{K^a}E & = &  R^{a\dagger}E + E R^a + E^a\,,
\\[5pt]
\label{app3-15082009-03} && E^a  \equiv (\Delta + 1) [E,x^a] + [E,x^b]M^{ab}
- \frac{1}{2}[[E,x^b],x^b]\partial^a\,,
\eeq
where $M^{ab}$ and $\Delta$ are given in \rf{080313-44xx},\rf{080313-44x}.
From \rf{app3-15082009-01}, \rf{delKAEbas01}, we see that the requirement of
invariance of the action under the conformal boost transformations amounts to
the equations
\be \label{app3-15082009-04}  R^{a \dagger}E + E R^a + E^a \approx 0\,.
\ee
In \rf{app3-15082009-04} and below, to simplify our formulas, we adopt the
following convention. Let $A$ be some operator. We use the relation $A
\approx 0$ in place of $\Phibr A\Phik=0$.

Using operator $E$ \rf{man03-15082009-02}-\rf{man03-15082009-11} and formula
\rf{app3-15082009-03}, we find immediately the operator $E^a$,
\be  \label{app3-15082009-05x} E^a = \left( \begin{array}{cc}
E_{\phi\phi\smone}^a + E_{\phi\phi\smzero}^a & \quad -  \upsilon^\ominussm
\bar\alpha^a
\\[7pt]
 - \alpha^a \bar\upsilon^\ominussm  & \quad  0
\end{array}\right)\,,
\ee
\beq
\label{app3-15082009-06x} && E_{\phi\phi\smone}^a = 2\Delta'\partial^a -(
\Delta' + N_\zeta)\alpha^a\albpar + (-\Delta' + N_\zeta )\alpar\bar\alpha^a
\,,
\\[7pt]
\label{app3-15082009-07x} && E_{\phi\phi\smzero}^a = (\Delta' - N_\zeta ) e_1
\bar\alpha^a + \alpha^a \eb_1 (\Delta' + N_\zeta )\,,
\eeq
where $e_1$, $\eb_1$ and $\Delta'$ are given in \rf{man03-15082009-09} and
\rf{080313-44x} respectively. Also, we note that the commutation relation
$[D,K^a]=K^a$ gives the following equation for the operator $R^a$:
\be \label{app3-15082009-05} [D,R^a] = R^a\,.\ee
Equations \rf{app3-15082009-04}, \rf{app3-15082009-05} constitute a complete
system of equations which allows us to determine the operator $R^a$ uniquely.
We now discuss the procedure of solving these equations.

Operator $E$ \rf{man03-15082009-02} is a second-order polynomial in the
derivative. From \rf{app3-15082009-05x}, we see that the operator $E^a$ is a
first-order polynomial in the derivative. The operator $R^a$ is also turned
out to be first-order polynomial in the derivative. Therefore it is
convenient to represent the operators $E$, $E^a$, and $R^a$ as power series
in the derivative,
\beq
\label{app3-15082009-06} && E = E_\smtwo + E_\smone + E_\smzero \,,
\\
\label{app3-15082009-07} && E^a = E_\smone^a + E_\smzero^a \,,
\qquad
R^a = R_\smone^a + R_\smzero^a \,,
\eeq
where the operators $E_{\smx{n}}$, $E_{\smx{n}}^a$, and $R_{\smx{n}}^a$ are
degree-$n$ homogeneous polynomials in the derivative. Explicit expressions
for the operators $E_{\smx{n}}$ and $E_{\smx{n}}^a$ can easily be read from
the respective expressions in \rf{man03-15082009-02}-\rf{man03-15082009-11}
and \rf{app3-15082009-05x}-\rf{app3-15082009-07x}. Using
\rf{app3-15082009-06}, \rf{app3-15082009-07} it is easy to see that
Eqs.\rf{app3-15082009-04} amount to the following equations:
\beq
\label{080314-01}
&& E_\smtwo R_\smone^a + h.c. \approx 0 \,,
\\[3pt]
\label{080314-02}
&& E_\smtwo R_\smzero^a +E_\smone R_\smone^a +  h.c. \approx 0 \,,
\\[3pt]
\label{080314-03}
&& \Bigl( E_\smone R_\smzero^a + E_\smzero R_\smone^a + h.c. \Bigr) +
E_\smone^a \approx 0 \,,
\\[3pt]
\label{080314-04}
&& \Bigl( E_\smzero R_\smzero^a + h.c. \Bigr)  + E_\smzero^a \approx 0 \,.
\eeq
We now present our procedure for solving Eqs.\rf{app3-15082009-05} and
\rf{080314-01}-\rf{080314-04}.

{\bf i}) We note that the most general operators $R_{\smx{n}}^a$
\rf{app3-15082009-07} acting on 2-vector $\Phik$ \rf{10122208-01} can be
presented as $2\times 2$ matrices given by
\be \label{22092009-01} R_{\smx{n}}^a = \left( \begin{array}{cc}
R_{\phi\phi\smx{n}}^a  &  R_{\phi t\smx{n}}^a
\\[7pt]
R_{t\phi\smx{n}}^a  &  R_{t t\smx{n}}^a
\end{array}\right)\,,\qquad n=0,1\,.
\ee
Requiring the operator $R^a$ to satisfy Eq.\rf{app3-15082009-05} and
constraints \rf{coco-01}-\rf{coco-04} we find the following expressions%
\footnote{ As a realization of the operator $R^a$ on the gauge field $\Phik$
is defined up to gauge transformation \rf{02112009man-09} we ignore
contributions to $R^a$ that can be removed by gauge transformation
\rf{02112009man-09}.}
\beq \label{22092009-01nn01}
&& R_{\phi\phi \smzero}^a =  r_{0,1}\bar\alpha^a  + \alpha^a \rb_{0,1}\,,
\\
&& R_{\phi\phi \smone}^a = r_{1,1}\partial^a + r_{1,5}\alpha^a \albpar +
r_{\epsilon 1}\epsilon_0^{ab}\partial^b\,,
\\[7pt]
&& R_{\phi t \smzero}^a = r_{0,4}\bar\alpha^a\,,
\\
&& R_{\phi t \smone}^a = 0\,,
\\[7pt]
&& R_{ t\phi \smzero}^a = r_{0,5}(1+\epsilon)\alpha^a\,,
\\
&& R_{t\phi \smone}^a = 0\,,
\\
\label{22092009-01nn02}
&& R_{t t\smx{n}}^a=0\,,\qquad n=0,1\,,
\eeq
where the operators $r_{0,1}$, $r_{0,2}$, $r_{0,4}$, $r_{0,5}$, $r_{1,1}$,
$r_{1,5}$ , $r_{\epsilon 1}$ independent of the oscillators $\alpha^a$ are
given by
\beq
&& r_{0,1} = \zeta  \rwt_{0,1} \bar\upsilon^\oplussm\,, \qquad \rb_{0,1} =
\upsilon^\oplussm \widetilde\rb_{0,1}\bar\zeta\,,
\hspace{1.2cm}
\\[6pt]
&& r_{1,1} = \upsilon^\oplussm \rwt_{1,1}\bar\upsilon^\oplussm\,,
\qquad
r_{\epsilon,1} = \upsilon^\oplussm \rwt_{\epsilon,1}\bar\upsilon^\oplussm\,,
\qquad
r_{1,5} = \upsilon^\oplussm \rwt_{1,5}\bar\upsilon^\oplussm\,,
\\[6pt]
\label{22092009-10} && r_{0,4} = \upsilon^\oplussm \rwt_{0,4}\,,
\hspace{1.3cm}
r_{0,5} =  \rwt_{0,5}\bar\upsilon^\oplussm\,,
\eeq
and the quantities $\rwt_{0,1}$, $\widetilde\rb_{0,1}$, $\rwt_{0,4}$,
$\rwt_{0,5}$, $\rwt_{1,1}$, $\rwt_{1,5}$ , $\rwt_{\epsilon 1}$ independent of
the oscillators $\alpha^a$, $\zeta$, $\upsilon^\oplussm$,
$\upsilon^\ominussm$ remain as undermined constants. Below, we  determine
these quantities by using Eqs.\rf{080314-01}-\rf{080314-04}.

Before analyzing Eqs.\rf{080314-01}-\rf{080314-04} we explain our
terminology. Introducing the notation $\XX$ for the left hand side of
Eqs.\rf{080314-01}-\rf{080314-04} we note that $\XX$  is a $2\times 2$ matrix
acting on 2-vector $\Phik$ \rf{10122208-01}. Using the notation
\be
\XX = \left( \begin{array}{cc}
\XX_{\phi\phi}  &  \XX_{\phi t}
\\
\XX_{t\phi}  &  \XX_{t t}
\end{array}\right)\,,
\ee we note that the $2\times 2$ matrix equation $\XX \approx 0$ amounts to
the four equations, $\phibr\XX_{\phi\phi}\phik=0$,  $ \phibr \XX_{\phi
t}\tk=0$, $\tbr \XX_{t\phi}\phik=0$, $ \tbr \XX_{t t}\tk =0$. We refer to
these four equations to as the respective $\phi\phi$-, $\phi t$-, $t\phi$-,
and $tt$-parts of the equation $\XX\approx 0$. We now turn to analysis of
Eqs.\rf{080314-01}-\rf{080314-04}.

{\bf ii}) Using $\phi\phi$ part of Eq.\rf{080314-01} and the relation
\be
\label{080314-08} E_{\phi\phi\smtwo } R_{\phi\phi\smone}^a=
r_{1,1}E_{\phi\phi\smtwo}
\partial^a + r_{1,5}\Box \alpha^a \albpar - r_{1,5} \alpar \albpar\partial^a
+ r_{\epsilon,1} \Box \epsilon_0^{ab}\partial^b \,, \ee
we find
\be \rwt_{1,1}^\dagger = \rwt_{1,1}\,,\qquad  \rwt_{\epsilon,1}^\dagger =
\rwt_{\epsilon,1}\,,\qquad \rwt_{1,5} = 0\,.\ee

{\bf iii}) Making use of $\phi\phi$ part of Eq.\rf{080314-02} and the
relations
\beq
\label{080314-10} && E_{\phi\phi\smtwo} R_{\phi\phi\smzero}^a = \alpha^a
E_{\phi\phi\smtwo} \rb_{0,1} + r_{0,1} E_{\phi\phi\smtwo} \bar\alpha^a -
\alpar \partial^a \rb_{0,1}\,,
\\[5pt]
\label{080314-11} && E_{\phi\phi\smone} R_{\phi\phi\smone}^a = e_1
r_{1,1}\albpar
\partial^a + \alpar \partial^a \eb_1 r_{1,1}\,,
\eeq
we obtain
\be \label{080314-13} \rb_{0,1} = [\eb_1,r_{1,1}]\,,\qquad \rwt_{0,1}^\dagger
= -\widetilde\rb_{0,1}\,.\ee

{\bf iv}) Using $\phi\phi$ part of Eq.\rf{080314-03} and the relations
\beq
\label{080314-14} && E_{\phi\phi\smone} R_{\phi\phi\smzero}^a =  e_1
\rb_{0,1}\partial^a - e_1 \rb_{0,1}\alpha^a \albpar + \eb_1 r_{0,1} \alpar
\bar\alpha^a \,,
\\[5pt]
\label{080314-15} && E_{\phi\phi\smzero}R_{\phi\phi\smone}^a = m_1
r_{1,1}\partial^a + m_1 r_{\epsilon,1} \epsilon_0^{ab}\partial^b\,,
\\[5pt]
\label{080314-16} && E_{\phi t} R_{t \phi \smzero}^a = \upsilon^\ominussm
r_{0,5}\partial^a - \upsilon^\ominussm r_{0,5} \alpha^a \albpar -
\upsilon^\ominussm r_{0,5} \epsilon_0^{ab}\partial^b\,,
\eeq
we find
\be \rwt_{0,5} = 1\,, \qquad \rwt_{\epsilon,1} = -1\,,\ \qquad \rwt_{0,1}
=1\,,\qquad \rwt_{1,1}=-1\,. \ee

{\bf v}) Using $\phi t$ part of Eq.\rf{080314-02} and the relation
\be E_{\phi\phi \smtwo }R_{\phi t \smzero}^a + (E_{t \phi}
R_{\phi\phi\smone}^a)^\dagger = (r_{0,4} - r_{1,1} \upsilon^\ominussm)\albpar
+ (r_{0,4} - r_{\epsilon,1}\upsilon^\ominussm )\epsilon_0^{ab} \partial^b
\albpar \ee
we obtain
\be \rwt_{0,4}=-1\,, \qquad \rwt_{0,4} = \rwt_{\epsilon,1}\,.\ee

{\bf vi}) Using $\rwt_{0,1}$, $\widetilde\rb_{0,1}$, $\rwt_{0,4}$,
$\rwt_{0,5}$, $\rwt_{1,1}$, $\rwt_{1,5}$ , $\rwt_{\epsilon 1}$ above-given,
we make sure that all the remaining equations in
\rf{080314-01}-\rf{080314-04} are satisfied automatically. We note that in
analysis of $tt$ part of Eq.\rf{080314-03} we use the identity
\be \tbr \alpha^a \albpar \tk =  \tbr \alpar\bar\alpha^a  \tk \,,\ee
which can be proved by using self-duality constraint \rf{man10112009-02}.

\appendix{ On-shell D.o.F of self-dual field in $4d$ }\label{app-dofder}

We analyze on-shell D.o.F of the conformal self-dual field in $4d$ with
Lagrangian \rf{080312-23}. To this end we use light-cone gauge. In light-cone
frame, the space-time coordinates $x^a$ are decomposed as $x^a= x^+,
x^-,x^i$, where the light-cone coordinates in $\pm$ directions are defined as
$x^\pm=(x^3 \pm x^0)/\sqrt{2}$ and $x^+$ is taken to be a light-cone time.
The $so(2)$ algebra vector indices take values $i,j =1,2$. We adopt the
conventions: $\partial^i=\partial_i\equiv\partial/\partial x^i$,
$\partial^\pm=\partial_\mp \equiv
\partial/\partial x^\mp$.

We are going to prove that on-shell D.o.F of the self-dual field are
described by two $so(2)$ algebra self-dual complex-valued vector fields
$\phi_{-1}^i$, $\phi_{1}^i$ and one complex-valued scalar field $\phi_0$,
\be  \label{app02-14082009-01} \phi_{-1}^i\,, \qquad \phi_{1}^i\,, \qquad
\phi_0\,,\ee
which satisfy the equations of motion
\be  \label{app02-14082009-02} \Box \phi_{-1}^i -  \phi_1^i = 0 \,, \qquad
\Box \phi_1^i=0\,,\qquad \Box \phi_0=0\,. \ee
The vector fields satisfy the $so(2)$ self-duality constraint:
\be \label{app02-14082009-03} \pi^{ij}\phi_{-1}^j  = 0 \,,\qquad
\pi^{ij}\phi_1^j = 0 \,. \ee
Here and below, we use the notation
\be \label{app02-13082009-12} \pi^{ij} \equiv \delta^{ij} + \tau
\epsilon^{ij}\,,\qquad \bar\pi^{ij} \equiv \delta^{ij} - \tau \epsilon^{ij}
\,,\ee
where $\delta^{ij}$ is the Kronecker delta, while $\epsilon^{ij}$ is the
Levi-Civita symbol normalized as $\epsilon^{12}=1$.

In order to find on-shell D.o.F we use equations of motion obtained from
Lagrangian \rf{080312-23} and self-duality constraint \rf{08122208-02},
\beq \label{app02-13082009-01}
&& \partial^a F^{ab}(\phi_1) = 0 \,,
\\
\label{app02-13082009-02} && \partial^bF^{ba}(\phi_{-1}) + \partial^b t^{ba}
- \phi_1^a - \partial^a \phi_0 = 0 \,,
\\
\label{app02-13082009-03} && \Box \phi_0 +\partial^a \phi_1^a = 0\,,
\\
\label{app02-13082009-04} && F^{ab}(\phi_1) = \frac{\tau }{2} \epsilon^{abce}
F^{ce}(\phi_1)\,,
\\
\label{app02-13082009-05} && t^{ab} = \frac{\tau }{2} \epsilon^{abce}
t^{ce}\,.
\eeq

Taking into account the light-cone frame decomposition of the vector and
tensor fields \rf{scafiecol01},
\be \phi_{-1}^a =\phi_{-1}^+,\, \phi_{-1}^-,\, \phi_{-1}^i\,,\qquad\quad
\phi_1^a =\phi_1^+,\, \phi_1^-,\, \phi_1^i\,,\qquad
t^{ab} =t^{+-}\,,\, t^{+i}\,,\, t^{-i}\,,\, t^{ij}\,, \ee
we note that some of gauge transformations given
\rf{080312-24}-\rf{080312-27} can be represented as
\beq
\label{app02-13082009-06} && \delta \phi_1^+ = \partial^+ \xi_0\,,
\\
\label{app02-13082009-07} && \delta \phi_{-1}^+ = \partial^+ \xi_{-2} -
\lambda^+\,,
\\
\label{app02-13082009-08} && \delta \phi_{-1}^i = \partial^i \xi_{-2} -
\lambda^i\,,
\\
\label{app02-13082009-09} && \delta t^{+-} =  \partial^+ \lambda^- -
\partial^- \lambda^+ + \tau \epsilon^{ij} \partial^i \lambda^j \,,
\\
\label{app02-13082009-10} && \delta t^{+i} =  \bar\pi^{ij}(\partial^+
\lambda^j -  \partial^j \lambda^+)\,.
\eeq
From \rf{app02-13082009-06}, we see that the field $\phi_1^+$ can be gauge
away by using $\xi_0$ gauge transformation. From \rf{app02-13082009-07},
\rf{app02-13082009-08}, we see that the fields $\phi_{-1}^+$ and $\pi^{ij}
\phi_{-1}^j$ can be gauged away by using the respective $\xi_{-2}$ and
$\pi^{ij}\lambda^j$ gauge transformations. From \rf{app02-13082009-09},
\rf{app02-13082009-10}, we see that the fields $t^{+-}$ and
$\bar\pi^{ij}t^{+j}$ can be gauged away by using the respective $\lambda^-$
and $\bar\pi^{ij}\lambda^j$ gauge transformations. To summarize, we can
impose the following gauge conditions:
\beq
\label{app02-13082009-13} && \phi_1^+ =  0 \,,
\\
\label{app02-13082009-14} && \phi_{-1}^+ = 0 \,,
\\
\label{app02-13082009-15} && \pi^{ij} \phi_{-1}^j = 0\,,
\\
\label{app02-13082009-16} && \bar\pi^{ij} t^{+j} = 0\,,
\\
\label{app02-13082009-17} && t^{+-} = 0 \,.
\eeq
Using gauge conditions \rf{app02-13082009-13}-\rf{app02-13082009-17}, one can
make sure that equations \rf{app02-13082009-01}-\rf{app02-13082009-05} amount
to the following equations
\beq
\label{app02-13082009-18} && \Box \phi_1^a  = 0\,,
\\
\label{app02-13082009-20} && \partial^a \phi_1^a =0\,,
\\
\label{app02-13082009-21} && \partial^a \phi_{-1}^a + \phi_0 =0\,.
\\
\label{080312-20} && \Box \phi_{-1}^a +\partial^b t^{ba} -\phi_1^a = 0 \,,
\\
\label{app02-13082009-19} && \Box \phi_0 = 0 \,,
\eeq
\beq
\label{app02-13082009-22} && \pi^{ij} \phi_1^j = 0 \,,
\\
\label{app02-13082009-23} &&  \pi^{ij} t^{+j} = 0 \,,
\\
\label{app02-13082009-24} && \bar\pi^{ij} t^{-j} = 0 \,,
\\
\label{app02-13082009-25} && t^{ij} = 0 \,.
\eeq
We now analyze gauge conditions \rf{app02-13082009-13}-\rf{app02-13082009-17}
and equations \rf{app02-13082009-18}-\rf{app02-13082009-25}.

{\bf i}) In view of \rf{app02-13082009-15}, \rf{app02-13082009-22}, we see
that the $so(2)$ algebra vector fields $\phi_{-1}^i$, $\phi_1^i$ are indeed
satisfy the self-duality constraint given in \rf{app02-14082009-03}.

{\bf ii}) Equation \rf{app02-13082009-18} leads to the 2nd equation in
\rf{app02-14082009-02}.

{\bf iii}) Differential constraints \rf{app02-13082009-20},
\rf{app02-13082009-21} and gauge conditions \rf{app02-13082009-13},
\rf{app02-13082009-14} tell us that the non-dynamical fields $\phi_1^-$ and
$\phi_{-1}^-$ can be expressed in terms of fields \rf{app02-14082009-01},
\be
\phi_1^- = - \frac{\partial^i}{\partial^+}\phi_1^i \,,
\qquad
\phi_{-1}^- = - \frac{\partial^i}{\partial^+}\phi_{-1}^i -
\frac{1}{\partial^+}\phi_0\,.
\ee

{\bf iii}) Equations \rf{app02-13082009-16}, \rf{app02-13082009-23} imply
\be \label{app02-13082009-26} t^{+i}= 0 \,.\ee

{\bf iv}) Taking into account \rf{app02-13082009-25}, \rf{app02-13082009-26}
and using Eq.\rf{080312-20} we obtain
\be \label{app02-13082009-27} \Box \phi_{-1}^i + \partial^+ t^{-i} -\phi_1^i
= 0 \,. \ee
Multiplying Eq.\rf{app02-13082009-27} by $\bar\pi^{ij}$ and using constraint
\rf{app02-13082009-15} we obtain the 1st equation in \rf{app02-14082009-02}.

{\bf v}) Multiplying Eq.\rf{app02-13082009-27} by $\pi^{ij}$ and taking into
account \rf{app02-13082009-15},\rf{app02-13082009-22} gives the equation
\be
\label{app02-13082009-29} \pi^{ij} t^{-j}=0\,.
\ee
Equations \rf{app02-13082009-24},\rf{app02-13082009-29} imply
\be \label{app02-13082009-30} t^{-i} = 0 \,.\ee
Taking into account \rf{app02-13082009-17}, \rf{app02-13082009-25},
\rf{app02-13082009-26}, \rf{app02-13082009-30}, we see that $t^{ab}=0$.

To summarize, we proved that on-shell D.o.F of the self-dual field with
Lagrangian \rf{080312-23} are described by fields given in
\rf{app02-14082009-01}. These fields satisfy equations of motion
\rf{app02-14082009-02} and self-duality constraints \rf{app02-14082009-03}.
Light-cone gauge Lagrangian which leads to equations of motion
\rf{app02-14082009-02} is given in \rf{man03-14082009-01}.

\appendix{  $\epsilon$-symbols }\label{app-epssymb}

In this Appendix, we describe various useful relations for $\epsilon$-symbols
we use in the paper.  We introduce the following $\epsilon$-symbols
constructed out of the Levi-Civita symbol $\epsilon^{a_1\ldots a_\nu b_1
\ldots b_\nu}$ and the oscillators:

\beq
&& \epsilon \equiv \frac{\tau}{(\nu!)^2}
\epsilon^{a_1\ldots a_\nu b_1 \ldots b_\nu}\alpha^{a_1}\ldots \alpha^{a_\nu}
\bar\alpha^{b_\nu}\ldots \bar\alpha^{b_1}\,,
\\[5pt]
\label{080313-42}  && \epsilon^a \equiv \frac{\tau}{\nu!(\nu-1)!}
\epsilon^{a_1\ldots a_\nu a b_2 \ldots b_\nu}\alpha^{a_1}\ldots
\alpha^{a_\nu} \bar\alpha^{b_\nu}\ldots \bar\alpha^{b_2}\,,
\\[5pt]
\label{080313-43}  && \bar\epsilon^a \equiv \frac{\tau}{\nu!(\nu-1)!}
\epsilon^{aa_2\ldots a_\nu b_1 \ldots b_\nu}\alpha^{a_2}\ldots \alpha^{a_\nu}
\bar\alpha^{b_\nu}\ldots \bar\alpha^{b_1}\,,
\\[5pt]
\label{080313-44}  && \epsilon_0^{ab} \equiv \frac{\tau}{((\nu-1)!)^2}
\epsilon^{a a_2\ldots a_\nu b b_2 \ldots b_\nu}\alpha^{a_2}\ldots
\alpha^{a_\nu} \bar\alpha^{b_\nu}\ldots \bar\alpha^{b_2}\,,
\\[5pt]
\label{080313-45}  && \epsilon^{ab} \equiv \frac{\tau}{\nu!(\nu-2)!}
\epsilon^{a_1\ldots a_\nu ab b_3 \ldots b_\nu}\alpha^{a_1}\ldots
\alpha^{a_\nu} \bar\alpha^{b_\nu}\ldots \bar\alpha^{b_3}\,,
\\[5pt]
\label{080313-46}  && \bar\epsilon^{ab} \equiv \frac{\tau}{\nu!(\nu-2)!}
\epsilon^{ab a_3\ldots a_\nu b_1 \ldots b_\nu}\alpha^{a_3}\ldots
\alpha^{a_\nu} \bar\alpha^{b_\nu}\ldots \bar\alpha^{b_1}\,,
\eeq
where $\tau$ is defined in \rf{man10112009-01}. We note the following helpful
relations for these $\epsilon$-symbols:
\beq
\label{080313-47}  && \epsilon^a \equiv [\epsilon,\alpha^a]\,,
\hspace{2cm}
\bar\epsilon^a \equiv [\bar\alpha^a,\epsilon]\,,
\\[5pt]
\label{080313-49}  && \epsilon^{ab} \equiv \{ [\epsilon,\alpha^a],\alpha^b
\}\,,
\qquad
\bar\epsilon^{ab} \equiv \{\bar\alpha^a,[\bar\alpha^b,\epsilon] \}\,,
\\[5pt]
\label{080313-51} && \epsilon_0^{ab} \equiv
\{[\bar\alpha^a,\epsilon],\alpha^b\}\,,
\nonumber\\[5pt]
&& \hspace{1cm} = \{\bar\alpha^a,\epsilon^b\} =
\{\bar\epsilon^a,\alpha^b\}\,.
\eeq
Our $\epsilon$-symbols satisfy the following hermitian conjugation rules:
\beq
\label{080313-52} \epsilon^\dagger = - \epsilon\,,\qquad
\epsilon^{a\dagger} = - \bar\epsilon^a\,,\qquad \epsilon_0^{ab\dagger} =
\epsilon_0^{ab}\,,\qquad \epsilon^{ab\dagger} =\bar\epsilon^{ab}\,.
\eeq
On space of ket-vector $\phik$ subject to the constraint
\be N_\alpha \phik= \nu\phik\,,\ee
we obtain the relation
\be \label{080313-54} \epsilon^2 \phik = \phik \,.\ee
It is this property of the $\epsilon$-symbol that is used for the definition
of the self-dual ket-vector $\tk$ \rf{man10112009-02}. One has the following
helpful identities involving $\epsilon$-symbols, the oscillators, and the
derivative
\beq
&&  \label{080313-55} \alpar \epsilon_0^{ab}\partial^b = \epsilon^b
\partial^b \partial^a -\epsilon^a \Box + \epsilon^{ab}\albpar\partial^b
\,,
\\[3pt]
&& \epsilon_0^{ab}\albpar \partial^b = -\bar\epsilon^b \partial^b\partial^a
+\bar\epsilon^a \Box +\alpar \bar\epsilon^{ab}\partial^b\,.
\eeq

\end{document}